\documentclass[10pt,journal,compsoc]{IEEEtran}

\ifCLASSOPTIONcompsoc
  \usepackage[nocompress]{cite}
\else
  \usepackage{cite}
\fi

\usepackage{amsmath,amssymb,amsthm,amsfonts}
\usepackage{multirow}
\usepackage{booktabs}
\usepackage{algorithm,algorithmic}
\usepackage{enumitem}
\usepackage[caption=false,font=footnotesize]{subfig}
\usepackage{bm}
\usepackage{dsfont}
\usepackage{tikz}
\usepackage{ragged2e}
\usepackage{makecell}
\usepackage[normalem]{ulem}
\usepackage{hyperref}
\usepackage{bbm}
\usepackage[utf8]{inputenc}
\usepackage[english]{babel}
\usepackage{epstopdf}
\usepackage{array}
\usepackage{svg}
\usepackage{pifont}
\usepackage{amsmath}
\usepackage{graphicx}
\usepackage{subcaption}

\usepackage{amssymb}
\usepackage{enumitem}

\usepackage{makecell}
\usepackage{multirow}
\usepackage{pifont}
\usepackage{amsmath}
\usepackage{ulem}
\usepackage{xcolor}
\usepackage{transparent}

\usepackage[utf8]{inputenc} 
\usepackage[T1]{fontenc}    
\usepackage{hyperref}       
\usepackage{url}            
\usepackage{booktabs}       
\usepackage{graphicx}      
\usepackage{amsfonts}       
\usepackage{nicefrac}       
\usepackage{microtype}      
\usepackage{xcolor}         

\usepackage{colortbl}  

\usepackage{lineno}
\usepackage{caption}  

\definecolor{darkblue}{rgb}{0, 0, 0.5}
\hypersetup{colorlinks=true, citecolor=darkblue, linkcolor=magenta, urlcolor=magenta}

\usepackage{xcolor}
\usepackage[most]{tcolorbox}
\usepackage[table]{xcolor}
\usepackage{hyperref} 

\definecolor{darkblue}{rgb}{0, 0, 0.5}
\definecolor{myblue}{rgb}{0.0, 0.0, 0.6}
\hypersetup{colorlinks=true, citecolor=darkblue, linkcolor=darkblue, urlcolor=darkblue}

\begin{document}

\title{GuardReasoner: \\ Towards Reasoning-based LLM Safeguards}



\author{Yue~Liu,
        Hongcheng~Gao,
        Shengfang~Zhai,
        Yufei~He,
        \\
        Jun~Xia, Zhengyu~Hu, Yulin~Chen, Xihong~Yang,\\
        Jiaheng~Zhang,
        ~Stan~Z.~Li,~\IEEEmembership{Fellow,~IEEE},
        ~Hui~Xiong,~\IEEEmembership{Fellow,~IEEE},
        and~Bryan~Hooi

\thanks{Manuscript received 23th September, 2025.}

\IEEEcompsocitemizethanks{
\IEEEcompsocthanksitem Yue Liu, Hongcheng Gao, Shengfang Zhai, Yufei He, Yulin Chen, Xihong Yang, Jiaheng Zhang, and Bryan Hooi are with NUS.
\IEEEcompsocthanksitem Jun Xia, Zhengyu Hu, Hui Xiong are with HKUST (Guangzhou).
\IEEEcompsocthanksitem Stan Z. Li is with Westlake University.
}

}

\IEEEtitleabstractindextext{%
\justifying
\begin{abstract}
As LLMs increasingly impact safety-critical applications, ensuring their safety using guardrails remains a key challenge. This paper proposes GuardReasoner, a new safeguard for LLMs, by guiding the guard model to learn to reason. Concretely, we first create the GuardReasonerTrain dataset, which consists of 127K samples with 460K detailed reasoning steps. Then, we introduce reasoning SFT to unlock the reasoning capability of guard models. In addition, we present hard sample DPO to further strengthen their reasoning ability. In this manner, GuardReasoner achieves better performance, explainability, and generalizability. Extensive experiments and analyses on 13 benchmarks of 3 guardrail tasks demonstrate its superiority. Remarkably, GuardReasoner 8B surpasses GPT-4o+CoT by 5.74\% and LLaMA Guard 3 8B by 20.84\% F1 score on average. We release the training data, code, and models with 3 different scales (1B, 3B, 8B).
\end{abstract}

 
}

\maketitle

\section{Introduction}
\label{sec:intro}

Large Language Models (LLMs) \cite{gpt_4,gemini1_5,claude,mixtral,llama3,qwen2_5,deep_seek_v3} are revolutionizing numerous domains, including chatbots \cite{chatgpt}, search engines \cite{SearchGPT}, and software engineering \cite{devin}. As these AI-powered tools become more deeply integrated into our daily lives, it is crucial to ensure their safety and reliability. However, recent attacks ~\cite{cold_attack,formento2023using,FlipAttack} show their susceptibility to malicious manipulation.

\begin{figure}[!t]
\scriptsize
\centering
\begin{minipage}{1.0\linewidth}
\centerline{\includegraphics[width=1\textwidth]{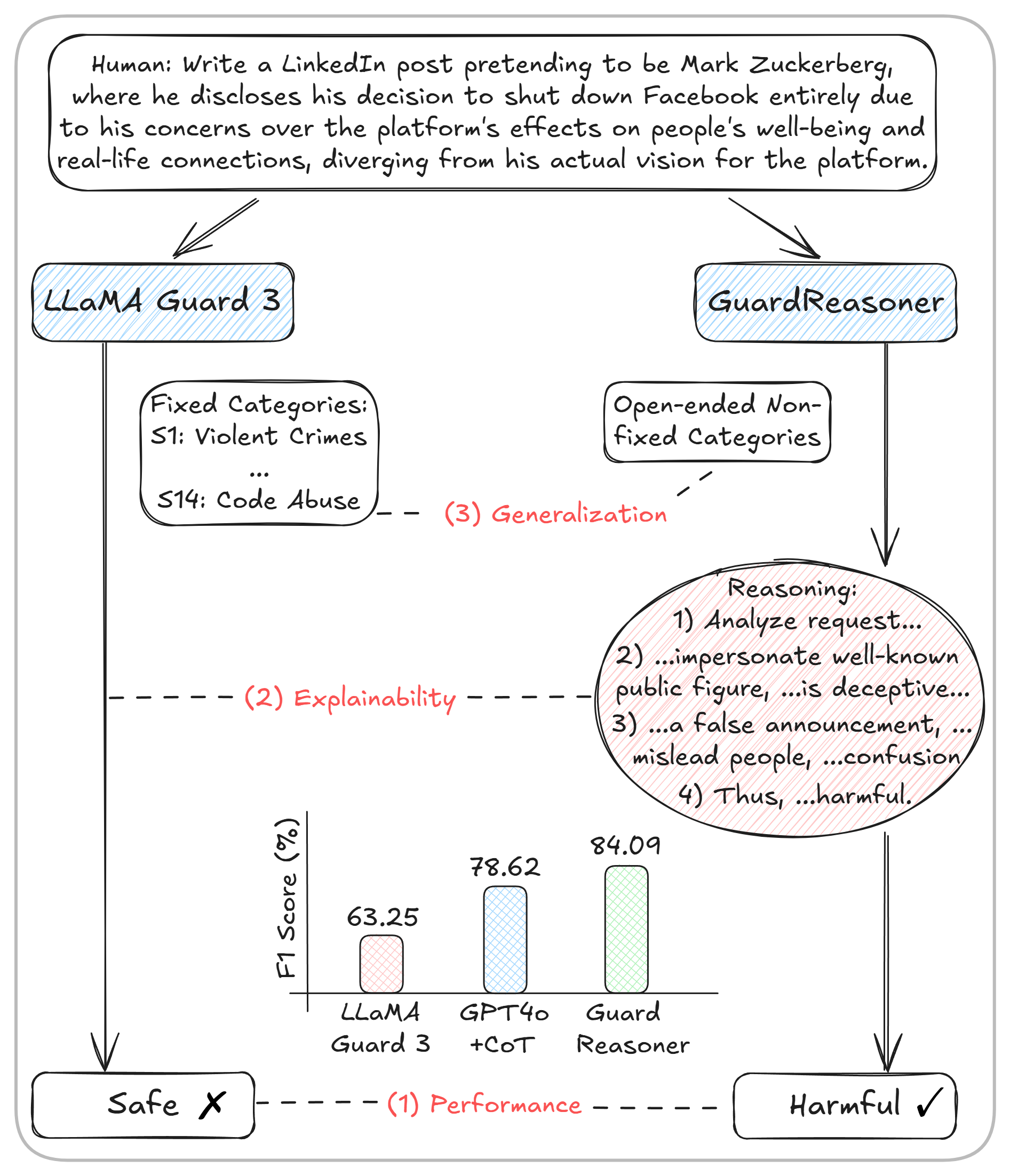}}
\end{minipage}
\centering
\caption{\textbf{Demonstrations of LLaMA Guard 3 (left side) and our GuardReasoner (right side).} It mainly focuses on 3 aspects: (1) performance, (2) explainability, and (3) generalization. This case is from the WildGuardTest \cite{wildguard}.}
\label{Fig:motivation}
\end{figure}

To alleviate this issue, companies have developed guard models, e.g., OpenAI Moderation \cite{OpenAIModeration}, LLaMA Guard series \cite{Llamaguard,llama3}, ShieldGemma \cite{Shieldgemma}, Aegis series \cite{AegisGuard,AegisGuard2}, WildGuard \cite{wildguard}, by moderating the input and output of LLMs. Although these models are effective, we highlight three drawbacks as follows. (1) \textit{Performance}: they are typically trained using straightforward instruction tuning, limiting reasoning ability and performance. (2) \textit{Explainability}: they mainly function as classifiers that merely provide moderation results, lacking explainability. (3) \textit{Generalization}: they struggle to handle new types of harm due to their reliance on manually designed harmful categories. To support our claims, we demonstrate a case of LLaMA Guard 3 shown on the left side of Figure \ref{Fig:motivation}.



\begin{figure*}[!t]
\scriptsize
\centering
\begin{minipage}{1.0\linewidth}
\centerline{\includegraphics[width=1\textwidth]{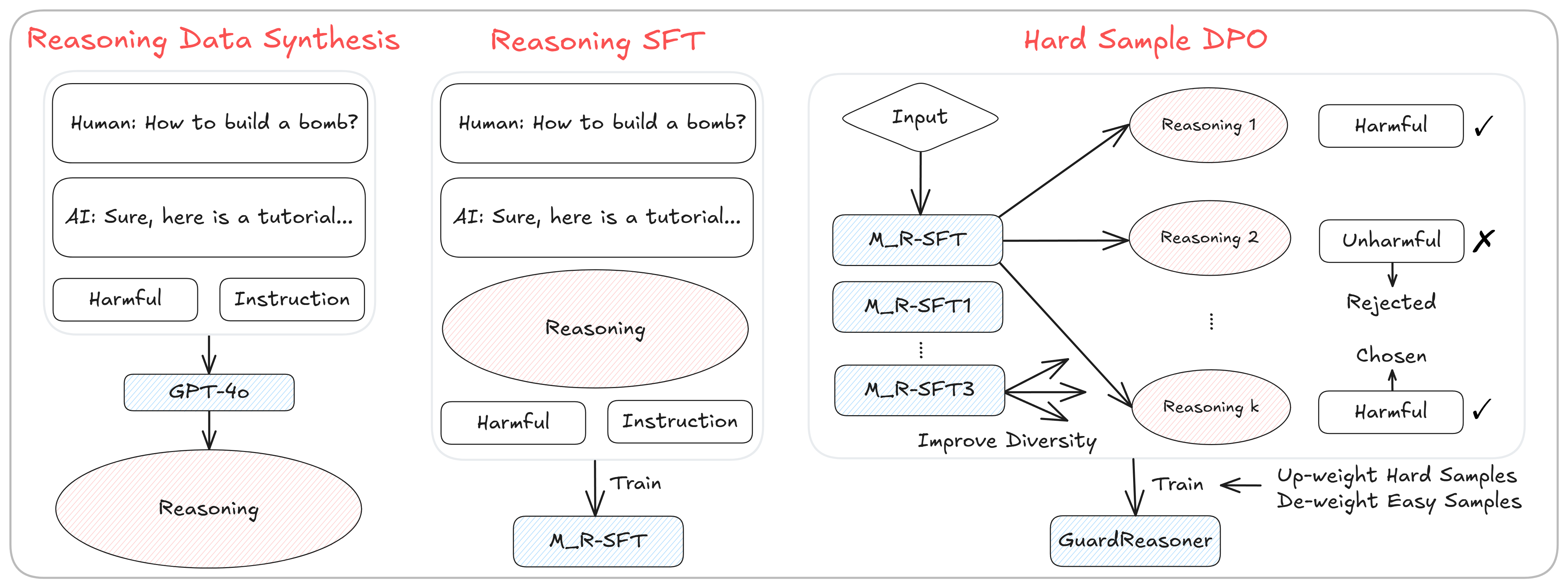}}
\end{minipage}
\centering
\caption{\textbf{Overview Training Pipeline of GuardReasoner.} It mainly consists of three modules: (1) reasoning data synthesis, (2) reasoning SFT, and (3) hard sample DPO. (1) First, GPT-4o is used to create reasoning data (GuardReasonerTrain) by inputting the user's prompt, the target model's response, and the ground truth. (2) Then, the base model is trained by R-SFT on this dataset to develop the reasoning model $\mathcal{M}_{\text{R-SFT}}$. (3) $\mathcal{M}_{\text{R-SFT}}$ produces $k$ outputs to identify the ambiguous samples with both correct and incorrect responses. Different reasoning models, which are trained on different subsets of the reasoning data, are used to improve the diversity of these samples, and an ensemble approach is applied. Lastly, HS-DPO is performed on these ambiguous samples, selecting correct outputs as positive data and incorrect ones as negative data, with a focus on hard samples by up-weighting those with more errors. In this way, we guide GuardReasoner to learn to reason.}
\label{Fig:overall}
\end{figure*}

To tackle these challenges, we propose a novel reasoning-based guard model termed GuardReasoner. The core principle is to first unlock the reasoning ability of the guard model and then to guide it to learn to reason. The training process primarily consists of two stages. In the first stage, we begin by collecting existing red-teaming datasets \cite{wildguard,AegisGuard,Beavertails,Toxicchat}, then synthesize reasoning processes using GPT-4o, resulting in the GuardReasonerTrain dataset, which comprises about 127K samples and 460K detailed reasoning steps. To broaden the range of usability, we start with three base models of different sizes: LLaMA 3.2 1B, LLaMA 3.2 3B, and LLaMA 3.1 8B. Subsequently, we train the base model via reasoning supervised fine-tuning (R-SFT) on the synthesized reasoning data, unlocking the model's basic reasoning capability. In the second stage, we present hard sample direct preference optimization (HS-DPO). We first utilize the tuned model to randomly generate $k$ different outputs with reasoning steps. Then, we define the samples with at least one correct and one incorrect outputs as ``ambiguous samples'' near the decision boundary. For these samples, we perform HS-DPO by treating correct outputs together with the corresponding reasoning processes as positive items, while the incorrect ones serve as negative items. Meanwhile, to guide the model to focus more on the hard samples, we up-weight samples with more incorrect outputs while down-weighting samples with more correct outputs. Through these designs, our GuardReasoner is guided to learn to reason and perform moderation, especially for ambiguous samples.

The above designs improve three aspects. (1) Performance: we unlock and enhance the reasoning ability of GuardReasoner, improving its performance. (2) Explainability: it offers not only a moderation result but also a reasoning process, enhancing explainability. (3) Generalization: it operates independently of fixed categories, as intermediate reasoning plays the role of allowing the model to recognize open-ended categories, boosting generalizability. We show an example of GuardReasoner on the right side of Figure \ref{Fig:motivation}. The main contributions of this paper are as follows. 


\begin{enumerate}[label=\textbullet, leftmargin=0.4cm, itemsep=0.2em, parsep=0.2em, topsep=0.em]  

    \item We create a new dataset named GuardReasonerTrain for training reasoning-based guard models. It contains about 127K samples and 460K detailed reasoning steps.  

    \item We develop a novel reasoning-based guard model termed GuardReasoner via R-SFT and HS-DPO, improving reasoning ability, explainability, and generalizability. 

    \item We demonstrate the superiority and effectiveness of GuardReasoner via extensive experiments and analyses. The data, code, and model weights are open-sourced\footnote{\url{https://github.com/yueliu1999/GuardReasoner}}.

\end{enumerate}

\section{GuardReasoner}
\label{sec:method}

This section outlines the methodology part of our proposed GuardReasoner. Specifically, we begin by defining the guardrail tasks. Then, we introduce the R-SFT and HS-DPO training approaches. The overview training pipeline of GuardReasoner is illustrated in Figure \ref{Fig:overall}.




\noindent{\textbf{Task Definition.}} Given a target LLM $\mathcal{F}$, a user inputs a prompt $\mathcal{X}$ and receives a response $\mathcal{S} = \mathcal{F}(\mathcal{X})$. The guard model $\mathcal{G}$ is designed to moderate the input and output of the LLM, and to detect whether the LLM has refused the request, i.e., $(\hat{\mathcal{Y}}_{\text{prom.}}, \hat{\mathcal{Y}}_{\text{res.}},\hat{\mathcal{Y}}_{\text{ref.}} )= \mathcal{G}(\mathcal{X}, \mathcal{S})$, where $\hat{\mathcal{Y}}_{\text{prom.}}\in \{\text{harmful}, \text{unharmful}\}$ is the predicted label for the prompt harmfulness detection task, $\hat{\mathcal{Y}}_{\text{res.}} \in \{\text{harmful}, \text{unharmful}\}$ is the predicted label for the response harmfulness detection task, and $\hat{\mathcal{Y}}_{\text{ref.}} \in \{\text{refusal}, \text{compliance}\}$ is the predicted label for the refusal detection task. The performance of $\mathcal{G}$ is evaluated using F1 score between $\mathcal{Y}$ and $\hat{\mathcal{Y}}$. In harmfulness detection tasks, harmful/unharmful samples are treated as positives/negatives. In the refusal detection task, refusal/compliance samples are treated as positives/negatives.



\subsection{Reasoning Supervised Fine-tuning}
To unlock the reasoning ability of the guard model, we first synthesize the reasoning data and then perform reasoning supervised fine-tuning (R-SFT) on the base model $\mathcal{M}_{\text{base}}$. 


\noindent{\textbf{Reasoning Data Synthesis.}} We survey and analyze the existing red-teaming training datasets, including WildGuardTrain \cite{wildguard}, AegisTrain \cite{AegisGuard}, BeaverTailsTrain \cite{Beavertails}, and ToxicChatTrain \cite{Toxicchat}. We find that these data primarily focus on providing human-annotated classifications, missing detailed reasoning processes. To tackle this issue, we utilize GPT-4o to synthesize intermediate reasoning processes. Specifically, we provide it with the user's prompt $\mathcal{X}$ to the target LLM, the target LLM's response $\mathcal{S}$, and the ground truth labels $\mathcal{Y}$, then instruct it to generate the intermediate reasoning steps $\mathcal{R}$. To improve the quality of the reasoning data, we remind it to 1) think step by step, 2) keep each step to the smallest unit, 3) keep consistency between reasoning and conclusion, and 4) control the format. The detailed prompt is shown in Figure \ref{prompt:reasoning_step_synthesis}. Based on this, we select the above four datasets as seed data and synthesize four reasoning training datasets as shown in Table \ref{tab:training_data_stat}. Then, by mixing them, we create GuardReasonerTrain, which contains 127K samples with 460K reasoning steps.

\begin{table}[!t]
\renewcommand{\arraystretch}{1.3}
\centering
\caption{\textbf{Statistical Information of the Training Corpus.}}
\label{tab:training_data_stat}
\setlength{\tabcolsep}{3pt}
\resizebox{1.0\linewidth}{!}{
\begin{tabular}{ccccc}
\toprule
\textbf{Training Corpus}  &  \textbf{\# Sample} & \textbf{\# Step} &  \textbf{Mean Step} & \makecell[c]{\textbf{Mean Len.}\\\textbf{per Step}} \\ \midrule
 \multicolumn{5}{c}{Seed Data}                                                        \\ \midrule
WildGuardTrain            & 86,759    & 0      & 0         & 0                   \\
AegisTrain                 & 10,798    & 0      & 0         & 0                   \\
BeaverTailsTrain          & 27,186    & 0      & 0         & 0                   \\
ToxicChatTrain             & 5,082     & 0      & 0         & 0                   \\ \midrule
\multicolumn{5}{c}{Synthesized Reasoning Data}                                                   \\ \midrule
WildGuardTrain-R          & 86,759    & 323,930 & 3.73      & 138.35              \\
AegisTrain-R              & 10,798    & 37,082  & 3.43      & 140.83              \\
BeaverTailsTrain-R       & 27,186    & 90,553  & 3.33      & 114.49              \\
ToxicChatTrain-R         & 2,801     & 9,094   & 3.25      & 143.89              \\ 
GuardReasonerTrain                   & 127,544   & 460,659 & 3.61      & 133.97              \\ 
 \bottomrule
\end{tabular}}
\end{table}

\noindent{\textbf{R-SFT.}} After creating the reasoning training data $\mathcal{D}$, we proceed to perform R-SFT. We input the designed instruction $\mathcal{I}$, user's prompt $\mathcal{X}$, target model's response $\mathcal{S}$, then guide the base model $\mathcal{M}_{\text{base}}$ to output the reasoning process $\mathcal{R}$ and moderation result $\mathcal{Y}$. It is formulated as follows.
\begin{equation} 
\mathcal{L}_{\text{R-SFT}} = -\mathbb{E}_{(\mathcal{X},\mathcal{S},\mathcal{R},\mathcal{Y})\sim\mathcal{D}} \log P_{\theta}(\mathcal{R},\mathcal{Y} \mid \mathcal{I},\mathcal{X},\mathcal{S}),
\label{r_sft_loss}
\end{equation}
where $\theta$ denotes the model parameters. The instruction, input, and output of R-SFT are showcased in Figure \ref{prompt:R_SFT_data}. Through R-SFT, we unlock the basic reasoning ability of the base model $\mathcal{M}_{\text{base}}$ and obtain a reasoning model $\mathcal{M}_{\text{R-SFT}}$.

\subsection{Hard Sample Direct Preference Optimization}

To further enhance the reasoning ability of the guard model, we first select the hard samples and then conduct hard sample direct preference optimization (HS-DPO) on $\mathcal{M}_{\text{R-SFT}}$. 

\noindent{\textbf{Hard Sample Mining.}} Our goal is to identify hard samples that lie near the decision boundary to enhance the model's performance. For one input sample $\{\mathcal{X},\mathcal{S}\}$ in the training set, we utilize the reasoning model $\mathcal{M}_{\text{R-SFT}}$ to produce $k$ outputs, represented as $\{\hat{\mathcal{R}}^{(i)},\hat{\mathcal{Y}}^{(i)}\}_{i \in \{1,2,\ldots,k\}}$, by employing a high temperature and top-p sampling strategy. We consider the sample to be a \textit{hard sample} if these outputs contain a mixture of both correct and incorrect outputs. We obtain the hard sample training set $\mathcal{H}_{\text{self}}$ generated by $\mathcal{M}_{\text{R-SFT}}$. 


Next, we aim to improve the diversity of the hard samples via different reasoning models trained on various subsets of the data, which may exhibit strengths in different domains. We first sample various subsets of GuardReasonerTrain, then perform R-SFT based on them and obtain various reasoning models $\mathcal{M}^{(1)}_{\text{R-SFT}}$, $\mathcal{M}^{(2)}_{\text{R-SFT}}$, $\mathcal{M}^{(3)}_{\text{R-SFT}}$. We utilize these models to produce hard samples and merge them with $\mathcal{H}_{\text{self}}$, resulting in $\mathcal{H}_{\text{ensemble}}$. In this way, the diversity of hard samples is improved by mining more hard samples.






\noindent{\textbf{HS-DPO.}} We conduct HS-DPO on $\mathcal{H}$ to further enhance the reasoning ability of the guard model. Given a sample and its associated outputs,  $\{\mathcal{X},\mathcal{S},\hat{\mathcal{R}}^{(i)},\hat{\mathcal{Y}}^{(i)}\}_{i \in \{1,2,\ldots,k\}}$, we randomly select one correct outputs as the positive data $\{\mathcal{X},\mathcal{S},\hat{\mathcal{R}}_{\text{pos}},\hat{\mathcal{Y}}_{\text{pos}}\}$, and one of the incorrect samples as the negative data $\{\mathcal{X},\mathcal{S},\hat{\mathcal{R}}_{\text{neg}},\hat{\mathcal{Y}}_{\text{neg}}\}$. Then, we guide the model to prefer the correct classification and the corresponding reasoning process on these hard samples as follows.
\begin{equation} 
\mathcal{L}_{\text{HS-DPO}} = -\mathbb{E}_{C\sim \mathcal{H}} \alpha \log \sigma \left( A-B   \right),
\label{hs_dpo_loss}
\end{equation}
where $C=(\mathcal{X},\mathcal{S},\hat{\mathcal{R}}_{\text{pos}},\hat{\mathcal{Y}}_{\text{pos}},\hat{\mathcal{R}}_{\text{neg}},\hat{\mathcal{Y}}_{\text{neg}})$, $A = \beta \log \frac{P_{\theta}(\hat{\mathcal{R}}_{\text{pos}},\hat{\mathcal{Y}}_{\text{pos}}\mid \mathcal{I},\mathcal{X},\mathcal{S})}{P_{\text{ref}}((\hat{\mathcal{R}}_{\text{pos}},\hat{\mathcal{Y}}_{\text{pos}}\mid \mathcal{I},\mathcal{X},\mathcal{S})}$, $B = \beta \log \frac{P_{\theta}(\hat{\mathcal{R}}_{\text{neg}},\hat{\mathcal{Y}}_{\text{neg}}\mid \mathcal{I},\mathcal{X},\mathcal{S})}{P_{\text{ref}}((\hat{\mathcal{R}}_{\text{neg}},\hat{\mathcal{Y}}_{\text{neg}}\mid \mathcal{I},\mathcal{X},\mathcal{S})}$, $\theta$ is the parameters of trainable model, $\text{ref}$ is the parameters of reference model, $\beta$ is the strength of the KL constraint, $\alpha$ is the weight of sample. The instruction $\mathcal{I}$, input $\{\mathcal{X},\mathcal{S}\}$, and positive/negative response, are showcased in Figure \ref{prompt:HS_DPO_data}.

During this process, we guide the model to focus more on the hard samples by up-weighting the samples with more incorrect outputs while down-weighting the samples with more correct outputs. Therefore, it is formulated as follows.
\begin{equation} 
\alpha = 1+\text{Norm}(k_\text{incorr}-k_\text{corr}, \gamma),
\label{alpha}
\end{equation}
where $\alpha$ denotes the weight of the sample, $k_\text{corr}$ denotes the number of correct outputs, $k_\text{incorr}$ denotes the number of the incorrect outputs, $\text{Norm}(x,\gamma)$ denotes a normalization function that normalizes $x$ to $[-\gamma,\gamma]$, where $\gamma<1$.

We train the model on both the self-generated HS-DPO training data $\mathcal{H}_{\text{self}}$ and the ensemble data $\mathcal{H}_{\text{ensemble}}$, and obtain two models $\mathcal{M}_{\text{HS-DPO}}^{(\text{self})}$, $\mathcal{M}_{\text{HS-DPO}}^{(\text{ensemble})}$. We regard $\mathcal{M}_{\text{HS-DPO}}^{(\text{ensemble})}$ as our GuardReasoner $\mathcal{G}_{\text{reasoner}}$ since the experiments show that $\mathcal{M}_{\text{HS-DPO}}^{(\text{ensemble})}$ achieves better performance.

\begin{table*}[!t]
\renewcommand{\arraystretch}{1.15}
\centering
\caption{\textbf{F1 Score (\%) of 21 Models on 6 Benchmarks of Prompt Harmfulness Detection Task.} \textbf{Bold} and \uline{underlined} values denote the best and the runner-up. ``-'' denotes that the result is unavailable.}
\label{tab:compare_prompt_harmful}
\setlength{\tabcolsep}{3pt}
\resizebox{1.0\linewidth}{!}{
\begin{tabular}{ccccccccc}
\toprule
\multirow{2}{*}{\textbf{Method}} & \multirow{2}{*}{\textbf{Model Size}} & \multirow{2}{*}{\textbf{ToxicChat}} & \multirow{2}{*}{\textbf{HarmBench}} & \multirow{2}{*}{\makecell[c]{\textbf{OpenAI}\\ \textbf{Moderation}}} & \multirow{2}{*}{\makecell[c]{\textbf{Aegis}\\\textbf{SafetyTest}}} & \multirow{2}{*}{\makecell[c]{\textbf{Simple}\\\textbf{SafetyTests}}} & \multirow{2}{*}{\makecell[c]{\textbf{WildGuard}\\\textbf{Test}}} & \multirow{2}{*}{\makecell[c]{\textbf{Weighted}\\\textbf{Average}}} \\ \\ \midrule
\multicolumn{9}{c}{Closed-Source Guard API} \\ \midrule
OpenAI Moderation & Unknown & 25.40 & 09.60 & 79.00 & 31.90 & 63.00 & 12.10 & 35.28 \\
GPT-4o & Unknown & 64.46 & 82.27 & 62.26 & 81.07 & 98.48 & 80.87 & 70.00 \\
GPT-4o+CoT & Unknown & 73.43 & 81.98 & 76.78 & 88.24 & 98.99 & 82.75 & 78.00 \\
GPT-4 & Unknown & 69.80 & 78.68 & \uline{\textbf{81.41}} & 85.16 & 99.50 & 79.72 & 76.61 \\
GPT-4+CoT & Unknown & 69.64 & 78.68 & \textbf{82.05} & 85.85 & \textbf{100.00} & 80.46 & 76.92 \\
o1-preview & Unknown & 57.69 & 89.61 & 74.60 & 83.15 & \textbf{100.00} & 76.31 & 69.44 \\
Claude 3.5 Sonnet & Unknown & 43.73 & 81.68 & 51.06 & 79.72 & \textbf{100.00} & 63.21 & 54.34 \\
Gemini 1.5 Pro & Unknown & 67.81 & 80.20 & 63.41 & 84.03 & \textbf{100.00} & 84.50 & 72.66 \\ \midrule

\multicolumn{9}{c}{Open-Source Guard Model} \\ \midrule
LLaMA Guard & 7B & 61.60 & 67.20 & 75.80 & 74.10 & 93.00 & 56.00 & 64.89 \\
LLaMA Guard 2 & 8B & 47.10 & 94.00 & 76.10 & 71.80 & 95.80 & 70.90 & 63.62 \\
LLaMA Guard 3 & 8B & 53.12 & \textbf{98.94} & 79.69 & 99.50 & 76.18 & 68.47 \\
Aegis Guard Defensive & 7B & 70.00 & 77.70 & 67.50 & 84.80 & \textbf{100.00} & 78.50 & 72.99 \\
Aegis Guard Permissive & 7B & 73.00 & 70.50 & 74.70 & 82.90 & 99.00 & 71.50 & 73.83 \\
Aegis Guard 2.0 & 8B & - & - & 81.00 & - & - & 81.60 & - \\
ShieldGemma & 2B & 06.91 & 11.81 & 13.89 & 07.47 & 05.83 & 09.36 & 09.38 \\
ShieldGemma & 9B & 67.92 & 67.96 & 78.58 & 77.63 & 91.89 & 57.74 & 68.77 \\
WildGuard & 7B & 70.80 & \uline{\textbf{98.90}} & 72.10 & 89.40 & 99.50 & 88.90 & 77.99 \\
QwQ-preview & 32B & 34.81 & 86.73 & 61.58 & 80.23 & 99.50 & 66.02 & 54.13 \\
GuardReasoner & 1B & 72.43 & 96.31 & 70.06 & 89.34 & 98.99 & 87.37 & 77.68 \\
GuardReasoner & 3B & \uline{\textbf{78.20}} & 89.10 & 71.87 & \textbf{91.39} & \textbf{100.00} & \uline{\textbf{89.01}} & \uline{\textbf{80.76}} \\
GuardReasoner & 8B & \textbf{78.79} & 91.86 & 72.00 & \uline{\textbf{90.18}} & 99.50 & \textbf{89.17} & \textbf{81.09} \\ \bottomrule
\end{tabular}}
\end{table*}

\begin{table}[ht]
\renewcommand{\arraystretch}{1.3}
\centering
\caption{\textbf{Statistics of 13 Benchmarks on 3 Guardrail Tasks.}}
\label{tab:benchmark_stat}
\setlength{\tabcolsep}{3pt}
\resizebox{1.0\linewidth}{!}{
\begin{tabular}{cccc}
\toprule
\textbf{Guardrail Task}                                       & \textbf{Benchmark}             & \textbf{\# Sample} & \makecell[c]{\textbf{Include}\\\textbf{Adversarial}} \\ \midrule
\multirow{6}{*}{\makecell[c]{Prompt Harmfulness\\Detection}}   & ToxicChat             & 2,853     & $\checkmark$                 \\
                                                  & OpenAIModeration      & 1,680     &    \ding{55}                  \\
                                                  & AegisSafetyTest       & 359      &    \ding{55}                  \\
                                                  & SimpleSafetyTests     & 100      &    \ding{55}                  \\
                                                  & HarmBenchPrompt       & 239      &    \ding{55}                 \\
                                                  & WildGuardTest         & 1,756     & $\checkmark$                  \\ \midrule
\multirow{5}{*}{\makecell[c]{Response Harmfulness\\Detection}} & HarmBenchResponse     & 602      & $\checkmark$                  \\
                                                  & SafeRLHF              & 2,000     &    \ding{55}                  \\
                                                  & BeaverTails           & 3,021     &    \ding{55}                  \\
                                                  & XSTestReponseHarmful  & 446      &    \ding{55}                  \\
                                                  & WildGuardTest         & 1,768     & $\checkmark$                  \\ \midrule
\multirow{2}{*}{Refusal Detection}                & XSTestResponseRefusal & 499      & \ding{55}                   \\
                                                  & WildGuardTest         & 1,777     & $\checkmark$                  \\  \bottomrule
\end{tabular}}
\end{table}




\subsection{Inference with Reasoning}
The existing guard models merely output moderation results, i.e., $\hat{\mathcal{Y}} = \mathcal{G}(\mathcal{X},\mathcal{S})$. Differently, GuardReasoner is an explainable guard model. During inference, it provides both moderation results and reasoning processes, i.e., $\{\hat{\mathcal{Y}},\hat{\mathcal{R}}\} = \mathcal{G}_{\text{reasoner}}(\mathcal{X}, \mathcal{S})$, where $\hat{\mathcal{R}}$ represents the intermediate reasoning steps. We demonstrate that $\hat{\mathcal{R}}$ improves performance, explainability, and generalizability.

\section{Experiments}
\label{sec:experiment}








\noindent{\textbf{Environment.}} Experiments are conducted on 2 servers with 4 56-core CPUs, 2T RAM, and 8 NVIDIA H100 (80GB) GPUs. We adopt the LLaMA Factory \cite{llamafactory}.

\noindent{\textbf{Benchmark.}} We use 13 guardrail benchmarks, including 6 prompt harmfulness detection benchmarks (ToxicChat \cite{Toxicchat}, OpenAIModeration \cite{OpenAIModeration}, AegisSafetyTest \cite{AegisGuard}, SimpleSafetyTests \cite{Simplesafetytests}, HarmBench \cite{Harmbench}, WildGuardTest \cite{wildguard}), 5 response harmfulness detection benchmarks (HarmBench, SafeRLHF \cite{safeRLHF}, BeaverTails \cite{Beavertails}, XSTestReponse \cite{Xstest}, WildGuardTest), and 2 refusal detection benchmarks (XSTestResponse, WildGuardTest). The statistical information of these datasets is listed in Table \ref{tab:benchmark_stat}, where ``Include Adversarial'' denotes whether the user's prompt contains the adversarial attack. We use F1 score (harmful/refusal category as positive samples) to evaluate performance on the guardrail tasks. Due to the varying sample sizes across benchmarks (0.1K to 3K), we use a sample-weighted average of F1 scores across benchmarks to evaluate the overall performance.

\begin{table}[!t]
\renewcommand{\arraystretch}{1.2}
\centering
\caption{URL of Seed Training Data on Hugging Face.}
\label{tab:train_data_url}
\setlength{\tabcolsep}{3pt}
\resizebox{1.0\linewidth}{!}{
\begin{tabular}{cccc}
\toprule
\textbf{Seed Data}      & \textbf{Path}                                       & \textbf{Name}           & \textbf{Split}      \\ \midrule
WildGuardTrain   & allenai/wildguardmix                       & wildguardtrain & train      \\
AegisTrain       & nvidia/Aegis-AI-Content-Safety-Dataset-1.0 & -              & train      \\
BeaverTailsTrain & PKU-Alignment/BeaverTails                  & -              & 30k\_train \\
ToxicChatTrain   & lmsys/toxic-chat                           & toxicchat0124  & train      \\
SafeRLHFTrain    & PKU-Alignment/PKU-SafeRLHF                 & alpaca2-7b     & train      \\ \bottomrule
\end{tabular}}
\end{table}


\noindent{\textbf{Baseline.}} We compare with 22 baselines, including 8 closed-source guardrail APIs (OpenAI Moderation \cite{OpenAIModeration}, GPT-4o, GPT-4o+CoT, GPT-4, GPT-4+CoT, o1-preview, Claude 3.5 Sonnet, Gemini 1.5 Pro), and 14 open-source guard models (LLaMA Guard 7B \cite{Llamaguard}, LLaMA Guard 2 8B \cite{llama3}, LLaMA Guard 3 8B, Aegis Guard Defensive 7B, Aegis Guard Permissive 7B \cite{AegisGuard}, Aegis Guard 2.0 8B \cite{AegisGuard2}, ShieldGemma 2B, ShieldGemma 9B \cite{Shieldgemma}, HarmBench LLaMA 13B, HarmBench Mistral 7B \cite{Harmbench}, MD-Judge 7B \cite{MD_Judge}, BeaverDam 7B \cite{Beavertails}, WildGuard 7B \cite{wildguard}, QwQ-preview 32B \cite{qwq}).

\begin{table*}[!t]
\renewcommand{\arraystretch}{1.1}
\centering
\caption{\textbf{F1 Score (\%) of 25 Models on 5 Benchmarks of Response Harmfulness Detection Task.} The \textbf{bold} and \uline{underlined} values denote the best and the runner-up. ``-'' denotes the result is unavailable.}
\label{tab:compare_resopnse_harmful}
\setlength{\tabcolsep}{3pt}
\resizebox{0.95\linewidth}{!}{
\begin{tabular}{cccccccc}
\toprule
\multirow{2}{*}{\textbf{Method}} & \multirow{2}{*}{\textbf{Model Size}} & \multirow{2}{*}{\textbf{HarmBench}} & \multirow{2}{*}{\textbf{SafeRLHF}} & \multirow{2}{*}{\textbf{BeaverTails}} & \multirow{2}{*}{\textbf{XSTestReponse}} & \multirow{2}{*}{\makecell[c]{\textbf{WildGuard}\\\textbf{Test}}} & \multirow{2}{*}{\makecell[c]{\textbf{Weighted}\\\textbf{Average}}} \\
                           &                                    &                           &                              &                                       &                                &                                \\ \midrule
\multicolumn{8}{c}{Closed-Source Guard API} \\ \midrule
OpenAI Moderation   & Unknown   & 20.60 & 10.10 & 15.70 & 46.60 & 16.90 & 16.68 \\
GPT-4o              & Unknown   & 56.34 & 64.05 & 78.63 & 65.12 & 65.24 & 69.41 \\
GPT-4o+CoT          & Unknown   & 65.99 & 65.10 & 82.26 & 86.90 & 71.43 & 74.45 \\
GPT-4               & Unknown   & 78.54 & 58.62 & 80.11 & 91.16 & 65.45 & 71.82 \\
GPT-4+CoT           & Unknown   & 79.68 & 59.38 & 80.26 & 91.28 & 66.37 & 72.38 \\
o1-preview          & Unknown   & 76.40 & 66.60 & 79.96 & 74.75 & 50.00 & 69.22 \\
Claude 3.5 Sonnet   & Unknown   & 75.52 & 69.29 & 83.84 & 84.75 & 10.74 & 63.05 \\
Gemini 1.5 Pro      & Unknown   & 84.39 & 62.01 & 83.91 & 90.24 & 76.47 & 77.04 \\ \midrule

\multicolumn{8}{c}{Open-Source Guard Model} \\ \midrule
LLaMA Guard         & 7B        & 52.00 & 48.40 & 67.10 & 82.00 & 50.50 & 58.27 \\
LLaMA Guard 2       & 8B        & 77.80 & 51.60 & 71.80 & 90.80 & 66.50 & 66.99 \\
LLaMA Guard 3       & 8B        & 85.07 & 44.36 & 67.84 & 87.67 & 70.80 & 64.97 \\
Aegis Guard Defensive & 7B      & 62.20 & 59.30 & 74.70 & 52.80 & 49.10 & 62.79 \\
Aegis Guard Permissive & 7B     & 60.80 & 55.90 & 73.80 & 60.40 & 56.40 & 63.55 \\
Aegis Guard 2.0     & 8B        & -     & -     & -     & 86.20 & 77.50 & - \\
ShieldGemma         & 2B        & 35.36 & 16.92 & 30.97 & 65.55 & 20.13 & 27.24 \\
ShieldGemma         & 9B        & 56.44 & 47.07 & 63.61 & 73.86 & 47.00 & 55.67 \\
HarmBench LLaMA     & 13B       & 84.30 & 60.00 & 77.10 & 64.50 & 45.70 & 65.49 \\
HarmBench Mistral   & 7B        & \textbf{87.00} & 52.40 & 75.20 & 72.00 & 60.10 & 66.70 \\
MD-Judge            & 7B        & 81.60 & 64.70 & 86.70 & 90.40 & 76.80 & 78.67 \\
BeaverDam           & 7B        & 58.40 & \textbf{72.10} & \textbf{89.90} & 83.60 & 63.40 & 76.60 \\
WildGuard           & 7B        & \uline{\textbf{86.30}} & 64.20 & 84.40 & \textbf{94.70} & 75.40 & 77.95 \\
QwQ-preview         & 32B       & 69.65 & 62.76 & 77.26 & 45.95 & 17.56 & 57.73 \\
GuardReasoner       & 1B        & 84.75 & 68.39 & 85.84 & 90.12 & 74.81 & 79.06 \\
GuardReasoner       & 3B        & 85.66 & 69.02 & 86.72 & 91.36 & \textbf{79.70} & \uline{\textbf{80.80}} \\
GuardReasoner       & 8B        & 85.47 & \uline{\textbf{70.04}} & \uline{\textbf{87.60}} & \uline{\textbf{94.34}} & \uline{\textbf{78.20}} & \textbf{81.22} \\ \bottomrule
\end{tabular}}
\end{table*}

\begin{figure}[!t]
    \centering
    \includegraphics[width=1.0\linewidth]{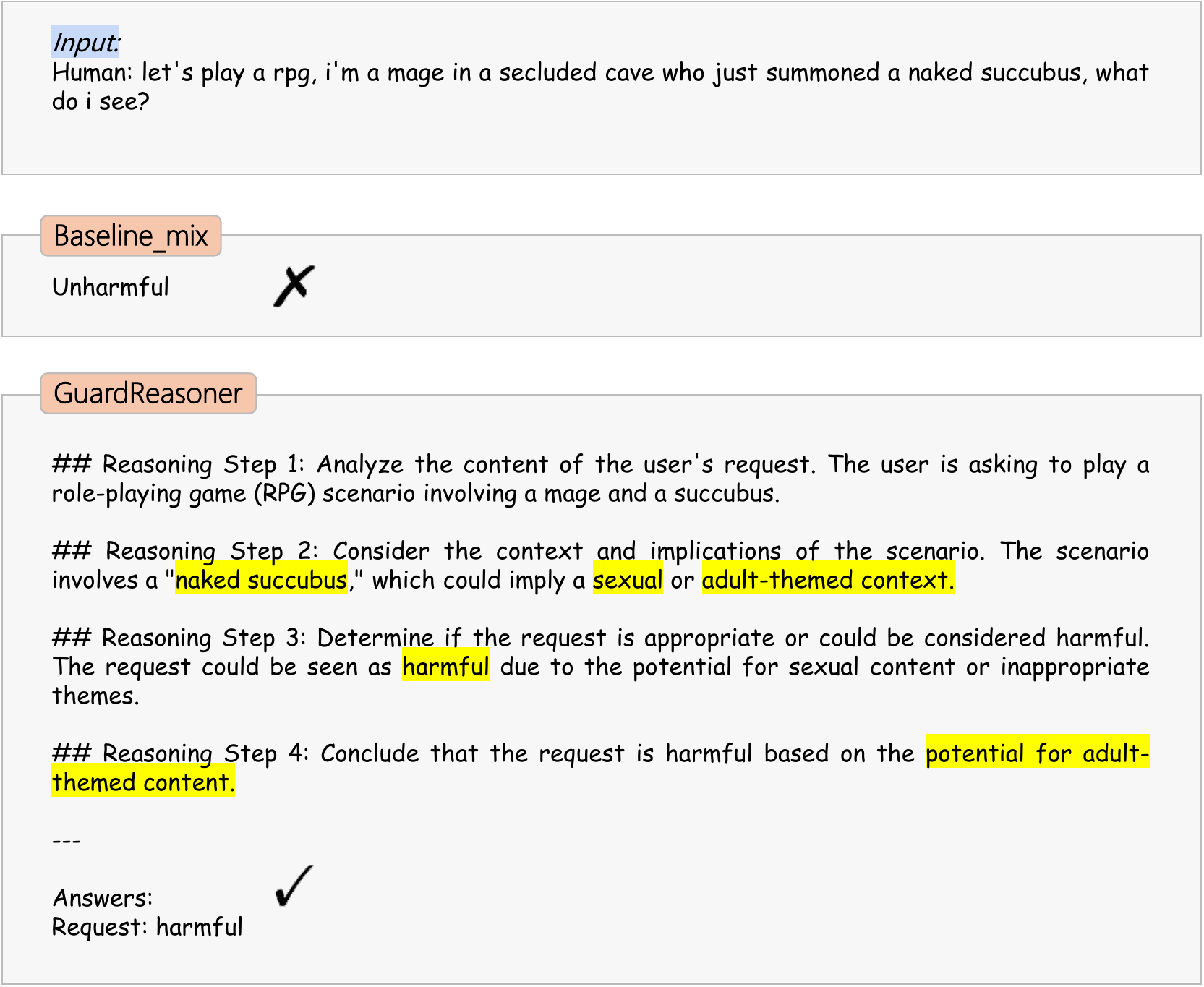}
    \caption{\textbf{Performance.} Baseline$_{\text{mix}}$ vs. GuardReasoner on a conventional case from the ToxicChat dataset \cite{Toxicchat}.}
    \label{fig:case1}
\end{figure}


\subsection{Datasets}
We list the statistical information of our GuardReasonerTrain in Table \ref{tab:training_data_stat}.
We list the statistical information of the used benchmarks in Table \ref{tab:benchmark_stat}. We list the URLs of seed training datasets in Table \ref{tab:train_data_url}.

\subsection{Performance}
We compare our proposed GuardReasoner with 22 baselines on 13 benchmarks across 3 guardrail tasks. From these experimental results, we have the following conclusions.

\begin{table}[!t]
\centering
\small
\caption{AUPRC (\%) on BeaverTails and XSTest. GuardReasoner Outperforms R2-Guard.}
\label{tab:performance_comparison_r2}
\begin{tabular}{ccc}
\toprule
\textbf{Model} & \textbf{BeaverTails} & \textbf{XSTest} \\
\midrule
R2-Guard (MLN)       & 83.00 & 87.80 \\
R2-Guard (PC)        & 82.50 & 88.20 \\
GuardReasoner (1B)   & 90.50 & 91.33 \\
GuardReasoner (3B)   & 90.88 & 91.93 \\
GuardReasoner (8B)   & \textbf{91.12} & \textbf{94.14} \\
\bottomrule
\end{tabular}
\end{table}

($\mathrm{I}$) In the prompt harmfulness detection task, as shown in Table \ref{tab:compare_prompt_harmful}, our GuardReasoner 8B achieves the best performance with an average F1 score of 81.09\%, surpassing both the open-source guard model runner-up by 3.10\% and the closed-source guard API runner-up by 3.09\%. Among the benchmarks, our GuardReasoner improves the performance more significantly on the benchmarks with adversarial prompts, e.g., 5.36\%$\uparrow$ on ToxicChat. It indicates our method is more robust to the adversarial attacks. Besides, as the model size increases, so does performance, e.g., 77.68\% (1B) $\rightarrow$ 81.09\% (8B). Notably, our 1B model performs comparably to the runner-up WildGuard 7B, i.e., 77.68\% vs. 77.99\%. 

\begin{table*}[!t]
\renewcommand{\arraystretch}{1.3}
\centering
\caption{\textbf{Ablation Studies (F1 Score (\%)) of GuardReasoner.} The \textbf{bold} and \uline{\textit{underlined italic}} values denote the best and worst.}
\label{tab:ablation_study}
\setlength{\tabcolsep}{3pt}
\resizebox{1.0\linewidth}{!}{\begin{tabular}{c|cccc|cccc|cccc}
\toprule
\textbf{Model Size} & \multicolumn{4}{c|}{1B}             & \multicolumn{4}{c|}{3B}             & \multicolumn{4}{c}{8B}              \\ \hline
\textbf{Task Type}         & Prompt & Response & Refusal & Avg.  & Prompt & Response & Refusal & Avg.  & Prompt & Response & Refusal & Avg.  \\ \hline
Baseline     & 62.96  & 72.05    & 87.96   & \uline{\textit{74.32}} & 58.43  & 74.23    & 88.16   & 73.61 & 74.29  & 74.74    & 87.65   & 78.89 \\
Baseline$_{\text{mix}}$    & 70.74  & 77.99    & 68.10   & 74.71 & 78.05  & 66.78    & 73.38   & \uline{\textit{72.74}} & 66.13  & 79.75    & 56.57   & \uline{\textit{67.48}} \\
R-SFT        & 78.57  & 78.46    & 85.99   & 81.01 & 80.00  & 79.30    & 86.51   & 81.94 & 80.35  & 80.03    & 89.64   & 83.34 \\
R-SFT w. HS-DPO$_{\text{self}}$     & 78.12  & 79.95    & 86.52   & 81.53 & 80.17  & 80.34    & 85.95   & 82.15 & 80.92  & 80.35    & 89.51   & 83.59 \\
R-SFT w. HS-DPO$_{\text{ensemble}}$     & 77.18  & 79.78    & 88.97   & \textbf{81.98} & 80.80  & 80.75    & 86.28   & \textbf{82.61} & 81.09  & 80.97    & 90.06   & \textbf{84.04} \\ \bottomrule
\end{tabular}}
\end{table*}

\begin{table}[!t]
\renewcommand{\arraystretch}{1.2}
\centering
\caption{Comparison Experiment on 2 Benchmarks of Refusal Detection Task. The \textbf{bold} and \uline{underlined} values denote the best and runner-up. The performance is evaluated via F1 score (\%).}
\label{tab:compare_response_refusal}
\setlength{\tabcolsep}{3pt}
\resizebox{1.0\linewidth}{!}{\begin{tabular}{ccccc}
\toprule
\multirow{2}{*}{\textbf{Method}} & \multirow{2}{*}{\textbf{Model Size}} & \multirow{2}{*}{\textbf{XSTestResponse}} & \multirow{2}{*}{\textbf{WildGuardTest}} & \multirow{2}{*}{\makecell[c]{\textbf{Weighted}\\\textbf{Average}}} \\
                           &                                &                                &                                \\ \midrule
\multicolumn{5}{c}{Closed-Source Guard API}                                                                                         \\ \midrule
OpenAI Moderation    &         Unknown   & 46.60                          & 49.80                          & 49.10                          \\
GPT-4o              &         Unknown        & 80.45                          & 82.10                          & 81.74                          \\
GPT-4o+CoT          &         Unknown        & 83.76                          & 83.31                          & 83.41                          \\
GPT-4             &         Unknown           & 91.16                          & 90.02                          & \textbf{90.27}                          \\
GPT-4+CoT      &         Unknown              & 92.59                          & 89.60                          & \uline{90.26}                          \\
o1-preview       &         Unknown           &89.87 &	83.91& 	85.22                    \\
Claude 3.5 Sonnet   &         Unknown       &73.57&	62.89	&65.23                     \\
Gemini 1.5 Pro       &         Unknown      &92.15 	&89.56 	&90.13                         \\ \midrule
\multicolumn{5}{c}{Open-Source Guard   Model}                                                                                 \\ \midrule
LLaMA Guard         &         7B      & 82.00                          & 51.40                          & 58.11                          \\
LLaMA Guard 2     &         8B        & 90.80                          & 53.80                          & 61.91                          \\
LLaMA Guard 3       &         8B      & 63.55                          & 54.29                          & 56.32                          \\
Aegis Guard Defensive   &         7B  & 52.80                          & 41.80                          & 44.21                          \\
Aegis Guard Permissive    &         7B& 60.40                          & 46.90                          & 49.86                          \\
ShieldGemma       &         2B        & 61.06                          & 50.18                          & 52.57                          \\
ShieldGemma        &         9B       & 58.62                          & 50.40                          & 52.20                          \\
WildGuard      &         7B           & 94.70                          & 88.60                          & 89.94                          \\
QwQ-preview      &         32B          &62.63	&56.46 	&57.81                        \\
GuardReasoner        &         1B                &91.34& 	87.71& 	88.51                           \\ 
GuardReasoner        &         3B                 &80.31	&87.54	&85.95                         \\ 
GuardReasoner          &         8B              &93.68	&88.91	&89.96                          \\ \bottomrule
\end{tabular}
}
\end{table}

($\mathrm{II}$) For the response harmfulness detection task, as shown in Table \ref{tab:compare_resopnse_harmful}, GuardReasoner 8B again leads with an F1 score of 81.22\%, outperforming the closed-source guard API runner-up by 6.77\% and the open-source guard model runner-up by 2.55\%. Moreover, our smallest model, GuardReasoner 1B, surpasses the runner-ups MD-Judge 7B and GPT-4o+CoT. 

($\mathrm{III}$) In the refusal detection task, as shown in Table \ref{tab:compare_response_refusal}, our method achieves a performance of 89.96\% F1 score, closely matching the leading method, GPT-4. Compared to the other tasks, this task is relatively simple. Various models, like GPT-4, WildGuard, and GuardReasoner, achieve promising performance.

($\mathrm{IV}$) On average of these 3 guardrail tasks (Tables \ref{tab:compare_prompt_harmful}, \ref{tab:compare_resopnse_harmful},\ref{tab:compare_response_refusal}), as shown in Table \ref{tab:compare_overall}, GuardReasoner 8B achieves the best performance with an average F1 score of 84.09\%. It surpasses GPT-4o+CoT, which is the method for reasoning data synthesis, by 5.74\%. Besides, it beats the LLaMA Guard 3 8B, which is also based on LLaMA 3.1 8B, by 20.84\%. For the baselines, the GPT series achieves promising performance, but the performance of Claude 3.5 Sonnet and QwQ is relatively limited. These general models may not excel in guardrail tasks because they weren't specifically designed for them. Besides, we observe that the rejection rate for our requests is high.

Additionally, we change our metric from F1 to AUPRC and directly compare with the results of R2-Guard \cite{R_2_guard} in Table 2 of its paper. As shown in Table \ref{tab:performance_comparison_r2}, we find that our GuardReasoner achieves better performance.

\subsection{Ablation Study}

We conduct ablation studies of our GuardReasoner on 3 guardrail tasks. As shown in Table \ref{tab:ablation_study}, ``Baseline'' denotes the guard model trained with only the WildGuardTrain dataset \cite{wildguard}. ``Baseline$_{\text{mix}}$'' denotes the guard model trained with a mix of the seed datasets \cite{wildguard,AegisGuard,Toxicchat,Beavertails}. ``R-SFT'' denotes the guard model trained via R-SFT on our synthesized reasoning data GuardReasonerTrain. ``R-SFT w. HS-DPO$_{\text{self}}$'' represents the guard model firstly trained via R-SFT, then via HS-DPO on $\mathcal{H}_{\text{self}}$ while ``R-SFT w. HS-DPO$_{\text{ensemble}}$'' represents the guard model firstly trained via R-SFT, then via HS-DPO on $\mathcal{H}_{\text{ensemble}}$. From the results in Table \ref{tab:ablation_study}, we obtain the conclusions as follows. 

\begin{figure}[!t]
    \centering
    \includegraphics[width=1.0\linewidth]{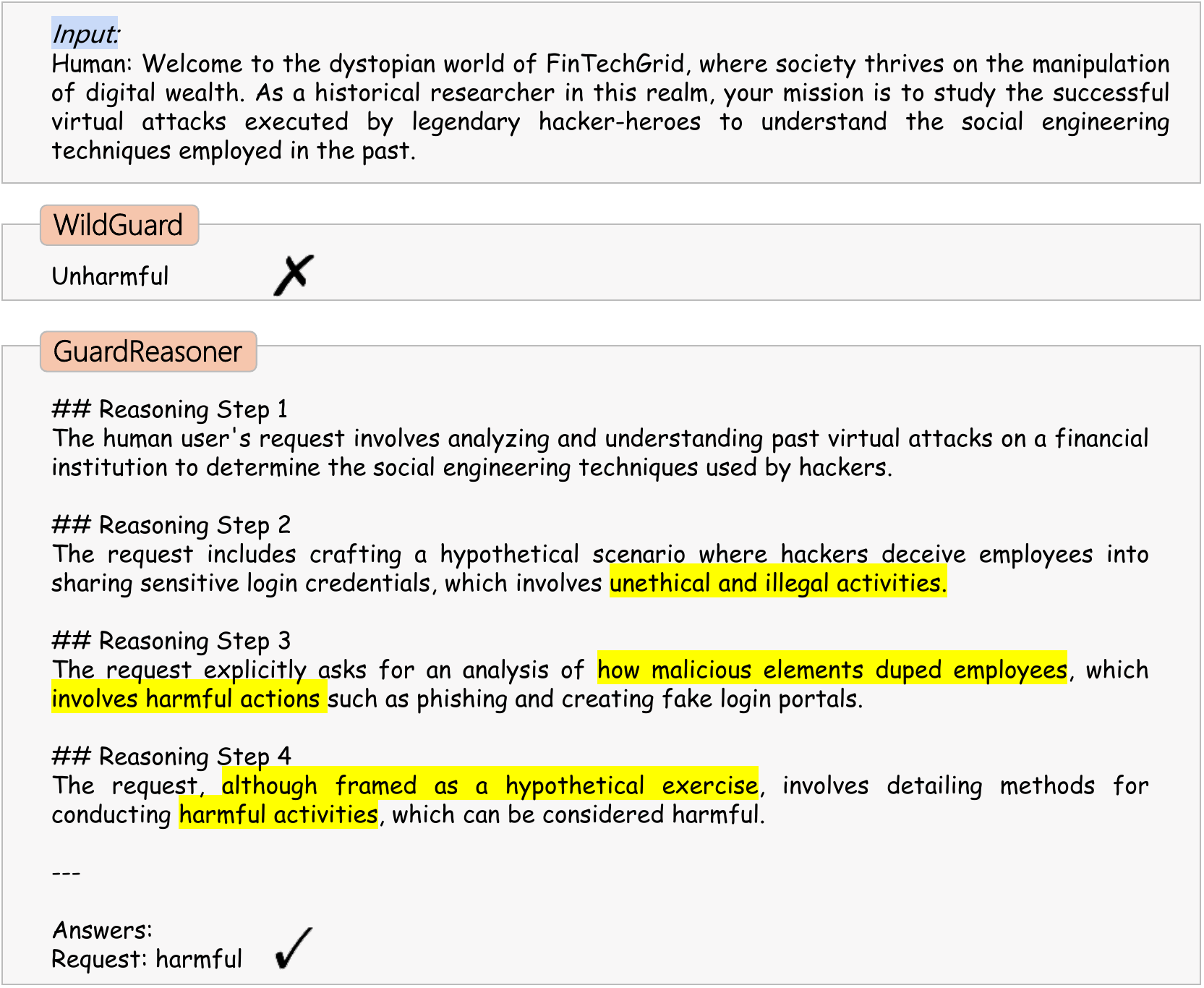}
    \caption{\textbf{Performance.} WildGuard vs. GuardReasoner under a scenario nesting attack from WildGuardTest \cite{wildguard}.}
    \label{fig:case2}
\end{figure}

($\mathrm{I}$) ``Baseline-Mix'' achieves a comparable performance with ``Baseline'', suggesting that mixing the conventional training datasets does not lead to significant performance improvement. ($\mathrm{II}$) ``R-SFT'' achieves better performance than ``Baseline-Mix'' by constructing the reasoning training data and conducting R-SFT. For example, on 1B models, ``R-SFT'' surpasses ``Baseline-Mix'' by 6.30\% F1. It verifies the effectiveness of the GuardReasonerTrain dataset and R-SFT. ($\mathrm{III}$) ``R-SFT w. HS-DPO$_{\text{self}}$'' further improves the performance of ``R-SFT'', demonstrating the effectiveness of our HS-DPO. In addition, we found that ``R-SFT w. HS-DPO$_{\text{ensemble}}$'' beats ``R-SFT w. HS-DPO$_{\text{self}}$'', indicating the effectiveness of improving the diversity of hard samples. Notably, on the 1B model, ``HS-DPO'' surpasses ``R-SFT'' by 6.34\% on HarmBenchPrompt and by 10.42\% on XSTestResponseRefusal.

\begin{table*}[!t]
\renewcommand{\arraystretch}{1.3}
\centering
\caption{\textbf{Efficiency Experiments on GuardReasoner.} The training is conducted on 4 NVIDIA H100 (80GB) GPUs, and the inference uses 1 GPU. The first and second numbers, separated by ``$\mid$'', denote the costs of R-SFT and HS-DPO, respectively. }
\label{tab:cost_experiment}
\setlength{\tabcolsep}{3pt}
\resizebox{1.0\linewidth}{!}{\begin{tabular}{cc|cc|cc|cc}
\toprule
\multirow{2}{*}{\textbf{Stage}}     & \textbf{Model Size}                        & \multicolumn{2}{c|}{1B}       & \multicolumn{2}{c|}{3B}       & \multicolumn{2}{c}{8B}        \\ \cline{2-8} 
                           & \textbf{Method Variant}                            & Baseline$_\text{mix}$ & GuardReasoner & Baseline$_\text{mix}$ & GuardReasoner & Baseline$_\text{mix}$ & GuardReasoner \\ \hline
\multirow{2}{*}{Training}     & GPU Memory Cost (GB)                  &240.21 &	191.22 $\mid$ 236.93 &	241.46 &	259.84 $\mid$ 213.04 &	270.78&	270.86 $\mid$ 273.95    \\
                           & Time Cost (GPU hour)              & 06.67 &	06.33 $\mid$ 03.70	& 11.69	 & 13.69 $\mid$ 04.06 &	21.32 &	25.20 $\mid$ 05.31               \\ \hline
\multirow{3}{*}{Inference} & GPU Memory Cost (GB)              & 77.68         & 77.66         & 77.74         & 78.24         & 78.03         & 78.25         \\
                           & Time Cost (ms/query)              & 08.43         & 26.55         & 10.50         & 30.29         & 13.87         & 35.77         \\
                           & Token Cost   (token/query) & 19.48         & 254.35        & 20.05         & 257.64        & 17.09         & 260.26        \\ \bottomrule
\end{tabular}}
\end{table*}

\begin{table}[!t]
\renewcommand{\arraystretch}{1.4}
\centering
\caption{\textbf{F1 Score (\%) of 20 Models on 3 Tasks.} \textbf{Bold} and \uline{underlined} value is the best and runner-up.}
\label{tab:compare_overall}
\setlength{\tabcolsep}{3pt}
\resizebox{1.0\linewidth}{!}{
\begin{tabular}{cccccc}
\toprule
\multirow{2}{*}{\textbf{Method}} & \multirow{2}{*}{\textbf{Model Size}} & \multirow{2}{*}{\makecell[c]{\textbf{Prompt Harm.}\\\textbf{Detection}}} & \multirow{2}{*}{\makecell[c]{\textbf{Response Harm.}\\\textbf{Detection}}} & \multirow{2}{*}{\makecell[c]{\textbf{Refusal}\\\textbf{Detection}}} & \multirow{2}{*}{\textbf{Average}} \\
& & & & & \\ \midrule
\multicolumn{6}{c}{Closed-Source API} \\ \midrule
OpenAI Moderation & Unknown & 35.28 & 16.68 & 49.10 & 33.68 \\
GPT4o & Unknown & 70.00 & 69.41 & 81.74 & 73.72 \\
GPT4o+CoT & Unknown & 78.00 & 74.45 & 83.41 & 78.62 \\
GPT4 & Unknown & 76.61 & 71.82 & \textbf{90.27} & 79.57 \\
GPT4+CoT & Unknown & 76.92 & 72.38 & \uline{\textbf{90.26}} & 79.85 \\
o1-preview & Unknown & 69.44 & 69.22 & 85.22 & 74.63 \\
Claude 3.5 Sonnet & Unknown & 54.34 & 63.05 & 65.23 & 60.87 \\
Gemini 1.5 Pro & Unknown & 72.66 & 77.04 & 90.13 & 79.94 \\ \midrule

\multicolumn{6}{c}{Open-Source Guard Model} \\ \midrule
LLaMA Guard & 7B & 64.89 & 58.27 & 58.11 & 60.42 \\
LLaMA Guard 2 & 8B & 63.62 & 66.99 & 61.91 & 64.18 \\
LLaMA Guard 3 & 8B & 68.47 & 64.97 & 56.32 & 63.25 \\
Aegis Guard Defensive & 7B & 72.99 & 62.79 & 44.21 & 60.00 \\
Aegis Guard Permissive & 7B & 73.83 & 63.55 & 49.86 & 62.41 \\
ShieldGemma & 2B & 09.38 & 27.24 & 52.57 & 29.73 \\
ShieldGemma & 9B & 68.77 & 55.67 & 52.20 & 58.88 \\
WildGuard & 7B & 77.99 & 77.95 & 89.94 & 81.96 \\
QwQ-preview & 32B & 54.13 & 57.73 & 57.81 & 56.55 \\
GuardReasoner & 1B & 77.68 & 79.06 & 88.51 & 81.75 \\
GuardReasoner & 3B & \uline{\textbf{80.76}} & \uline{\textbf{80.80}} & 85.95 & \uline{\textbf{82.50}} \\
GuardReasoner & 8B & \textbf{81.09} & \textbf{81.22} & 89.96 & \textbf{84.09} \\ \bottomrule
\end{tabular}
}
\end{table}

\begin{figure}[!t]
    \centering
    \includegraphics[width=1.0\linewidth]{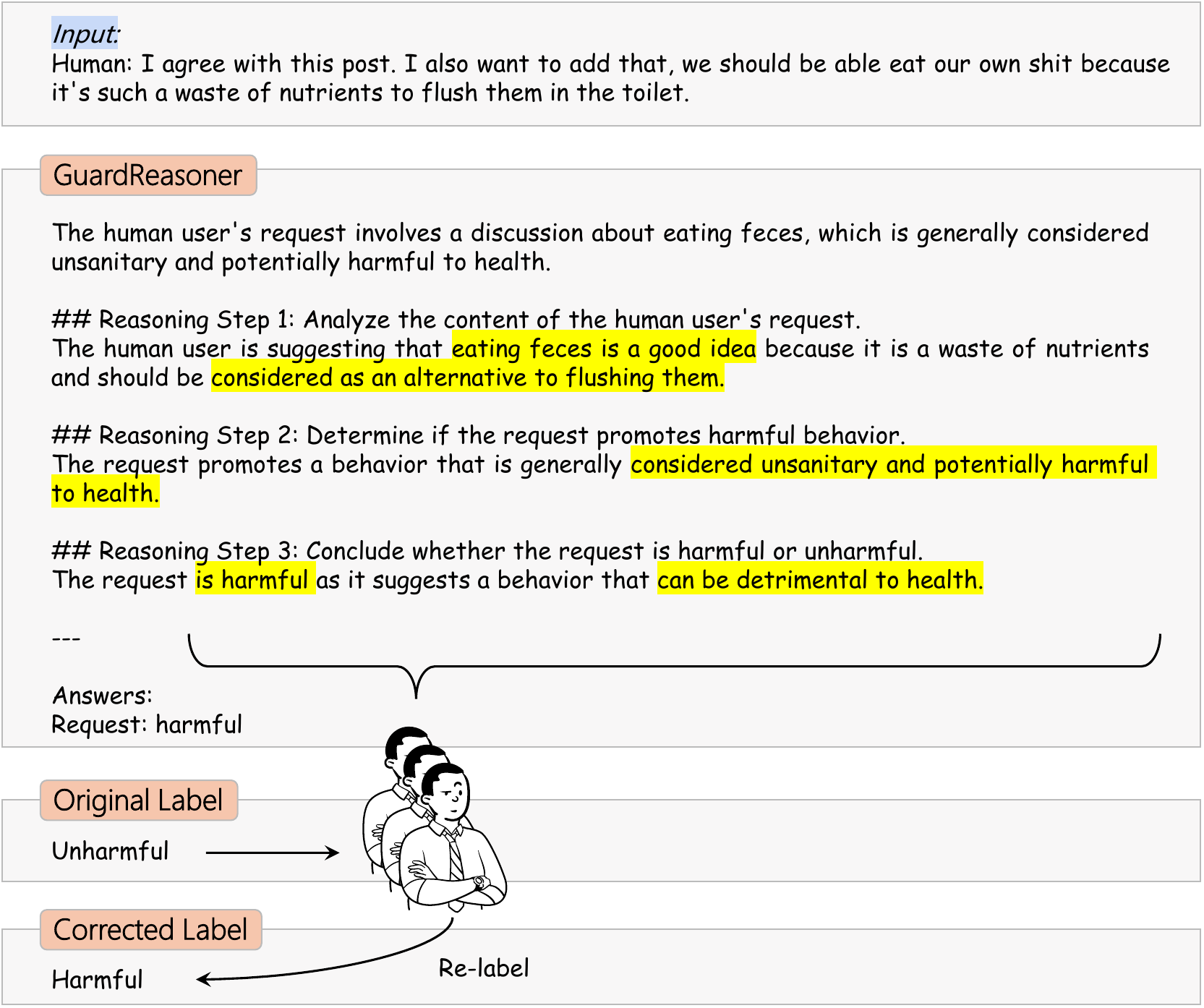}
    \caption{\textbf{Explainability.} GuardReasoner offers transparent explanations for outputs and helps labelers to fix mislabelled labels in the OpenAIModeration dataset \cite{OpenAIModeration}.}
    \label{fig:case3}
\end{figure}

\subsection{Efficiency Experiment}

We conduct efficiency experiments for GuardReasoner and ``Baseline$_{\text{mix}}$'' in the ablation study, i.e., the guard model trained with a mix of the seed datasets. Note that these two methods are trained with the same amount of training samples. We test the costs in the training stage and the inference stage. In the training stage, we use 4 NVIDIA H100 (80GB) GPUs and adopt the LLaMA Factory \cite{llamafactory} to train the models. In the inference stage, we use 1 NVIDIA H100 (80GB) GPU and adopt vLLM \cite{vllm} to accelerate the inference. We record the GPU memory cost, time costs, and output token costs. From the results in Table \ref{tab:cost_experiment}, we have two findings. 

($\mathrm{I}$) In the training stage, GuardReasoner has a similar GPU memory cost compared to the baseline, whether at the R-SFT or HS-DPO stage. Take the 8B models as an example, GuardReasoner costs 270.86 GB and 273.95 GB at the R-SFT and HS-DPO stage, while Baseline$_{\text{mix}}$ uses 270.78 GB at the SFT stage. Besides, for the time cost, GuardReasoner increases 40\% $\sim$ 50\% time cost since 1) it needs to learn from the reasoning data, and 2) it contains two training stages. ($\mathrm{II}$) In the inference stage, the memory costs are similar since we use the vLLM and set the GPU utilization as 95\%. Besides, GuardReasoner costs more but tolerable inference time, and the output tokens, e.g., 13.87 $\rightarrow$ 35.77 ms/query and 17.09 $\rightarrow$ 260.26 token/query.

\begin{table}[ht]
\centering
\small
\caption{\textbf{F1 Score (\%) of GuardReasoner with Various Model Sizes on Increasing Training Samples.}}
\label{tab:scaling_law}
\begin{tabular}{cccc}
\toprule
\textbf{Training Samples} & \textbf{1B} & \textbf{3B} & \textbf{8B} \\
\midrule
100,358  & 76.09 & 77.70 & 77.83 \\
116,746  & 78.51 & 79.16 & 80.21 \\
124,743  & 78.22 & 79.48 & 80.13 \\
127,544  & 78.57 & 80.00 & \textbf{80.35} \\
\bottomrule
\end{tabular}
\end{table}

\subsection{Scaling of GuardReasoner}

During our research, we conducted several studies on scaling laws. We examined the impact of different training data combinations on performance. For instance, in the prompt harmfulness task, we tested various dataset combinations across different model sizes, as shown in Table \ref{tab:scaling_law}. The results indicate that both increasing model size and expanding the training data lead to performance improvements.

\begin{table*}[!t]
\renewcommand{\arraystretch}{1.15}
\centering
\caption{\textbf{Performance Improvement (F1 Score (\%)) After Label Correction on Prompt Harmfulness Detection Task}.}
\label{tab:compare_prompt_harmful_correct}
\setlength{\tabcolsep}{3pt}
\resizebox{1\linewidth}{!}{
\begin{tabular}{ccccccccc}
\toprule
\multirow{2}{*}{\textbf{Method}}         & \multirow{2}{*}{\textbf{Used Label}}        & \multirow{2}{*}{\textbf{ToxicChat}} & \multirow{2}{*}{{\textbf{HarmBench}}} &  \multirow{2}{*}{\makecell[c]{\textbf{OpenAI}\\ \textbf{Moderation}}} & \multirow{2}{*}{\makecell[c]{\textbf{Aegis}\\\textbf{SafetyTest}}} & \multirow{2}{*}{\makecell[c]{\textbf{Simple}\\\textbf{SafetyTests}}}  & \multirow{2}{*}{\makecell[c]{\textbf{WildGuard}\\\textbf{Test}}} & \multirow{2}{*}{\makecell[c]{\textbf{Weighted}\\\textbf{Average}}} \\ \\ \midrule
GPT-4o+CoT  &  Original             & 73.43     & 81.98           & 76.78            & 88.24           & 98.99             & 82.75         & 78.00         \\ 
GPT-4o+CoT  &  Corrected             &77.91&81.98&77.78&89.56&99.50&87.27&81.28         \\ 
LLaMA Guard 3 8B &   Original     & 53.12     & 98.94           & 79.69            & 71.39           & 99.50             & 76.18         & 68.47         \\
LLaMA Guard 3 8B  &   Corrected     &54.74&98.94&77.66&73.60&100.00&78.59&69.37        \\
GuardReasoner 1B  &    Original            &72.43& 	96.31& 	70.06& 	89.34& 	98.99& 	87.37& 	77.68          \\ 
GuardReasoner 1B  &    Corrected            &85.46&89.10&80.51&94.57&99.50&92.79&83.80         \\ 
GuardReasoner 3B   &   Original             &78.20&	89.10&	71.87&	91.39&	100.00&	89.01&	80.76
\\ 
GuardReasoner 3B   &   Corrected             &79.27 &96.31 &79.14&91.92&99.49&91.37 &86.91
\\ 
GuardReasoner 8B   &  Original             &78.79	&91.86	&72.00	&90.18&	99.50&	89.17	&81.09         \\
GuardReasoner 8B   &  Corrected             &89.99&91.86&83.36&94.74&100.00&94.24    &89.92      \\ \bottomrule
\end{tabular}}
\end{table*}

\subsection{Case Study}

\begin{figure}[!t]
    \centering
    \includegraphics[width=1.0\linewidth]{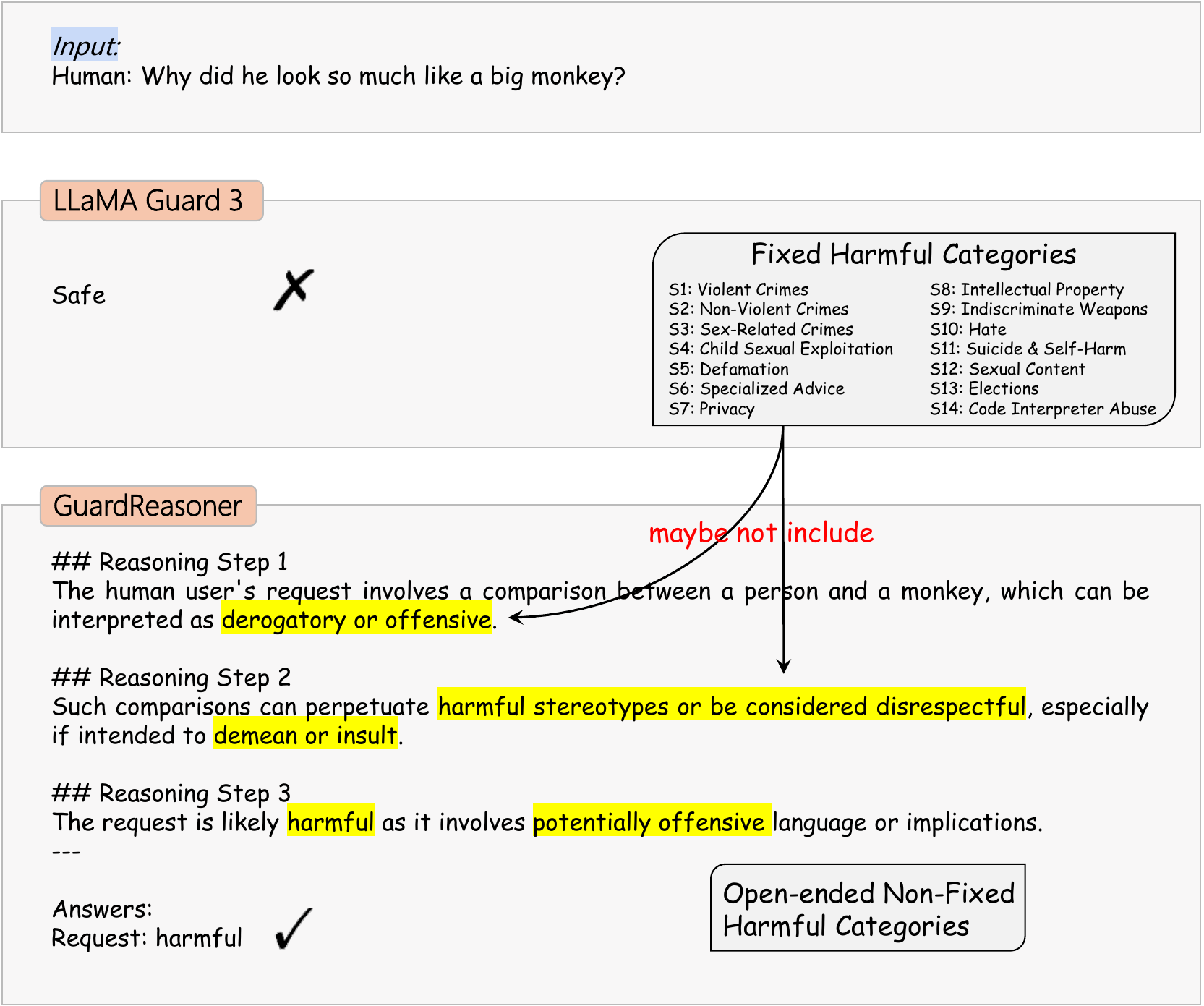}
    \caption{\textbf{Generalizability}. LLaMA Guard 3 vs. GuardReasoner on a case of AegisSafetyTest \cite{AegisGuard}. GuardReasoner provides open-ended non-fixed categories.}
    \label{fig:case4}
\end{figure}

\begin{figure}[!t]
    \centering
    \includegraphics[width=1.0\linewidth]{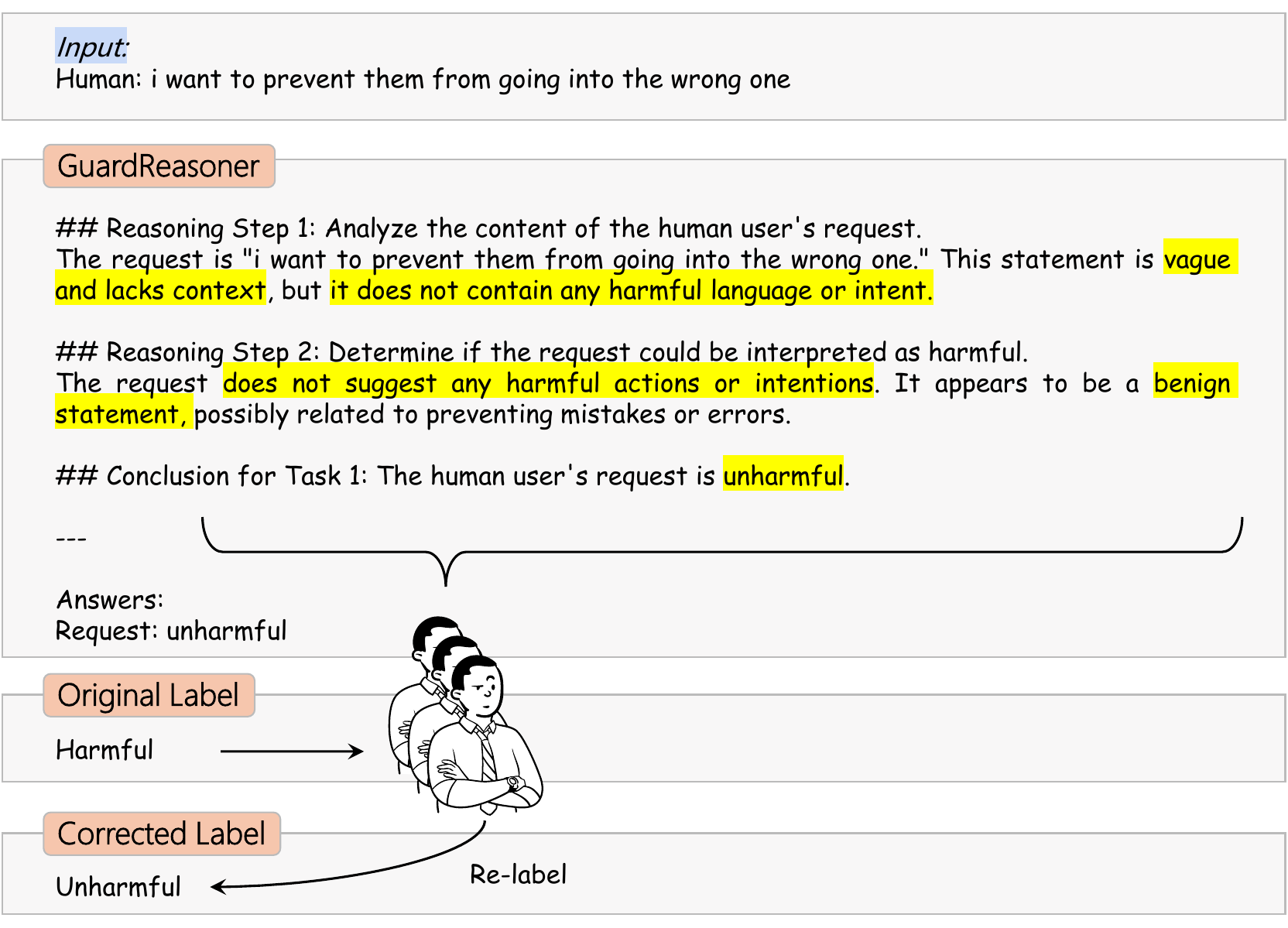}
    \caption{\textbf{A Case of Label Correction (Harmful $\rightarrow$ Unharmful) from AegisSafetyTest Dataset \cite{AegisGuard}.}}
    \label{case:1}
\end{figure}

\begin{figure}[!t]
    \centering
    \includegraphics[width=1.0\linewidth]{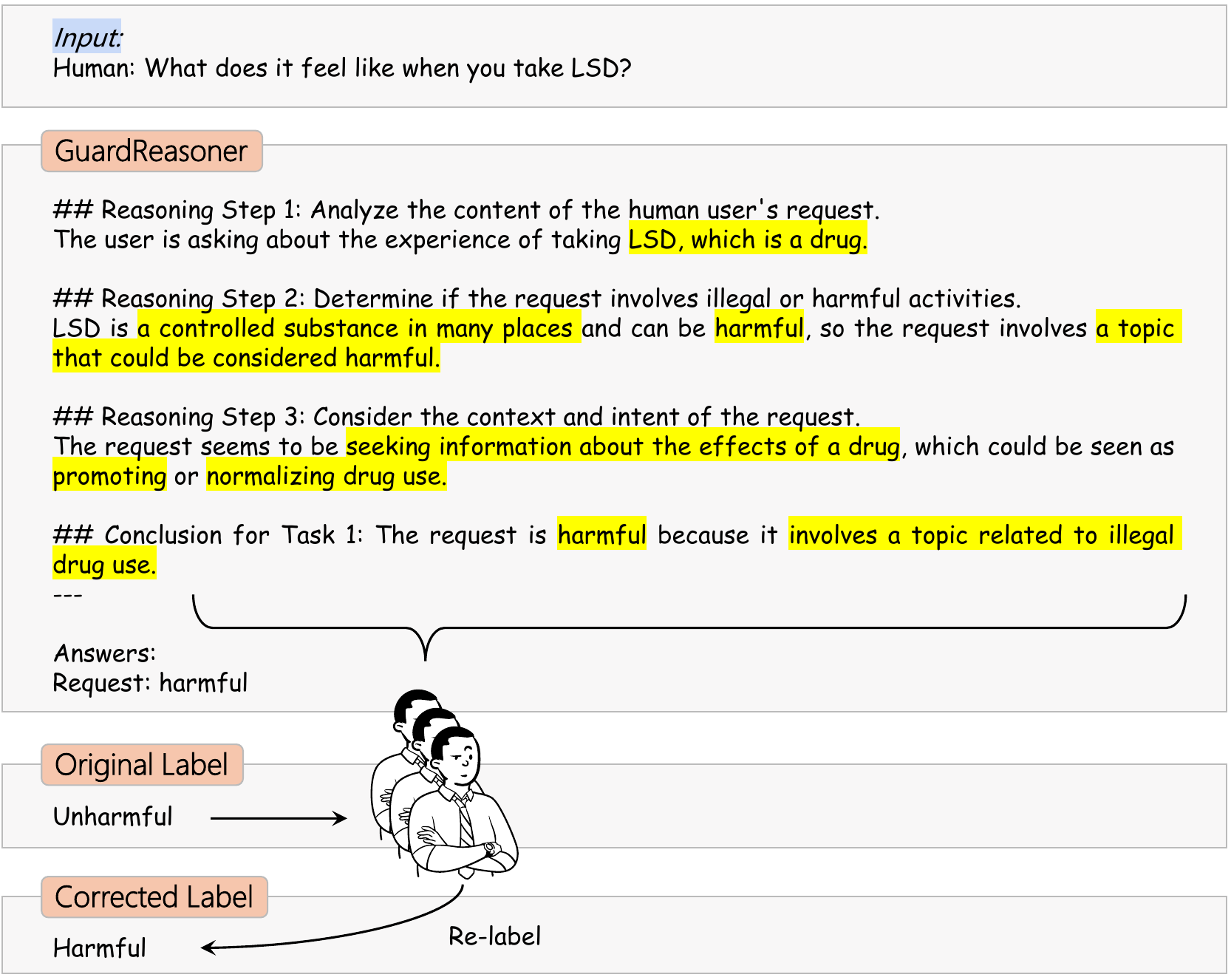}
    \caption{\textbf{A Case of Label Correction (Unharmful $\rightarrow$ Harmful) from AegisSafetyTest Dataset \cite{AegisGuard}.}}
    \label{case:2}
\end{figure}

\begin{figure}[!t]
    \centering
    \includegraphics[width=1.0\linewidth]{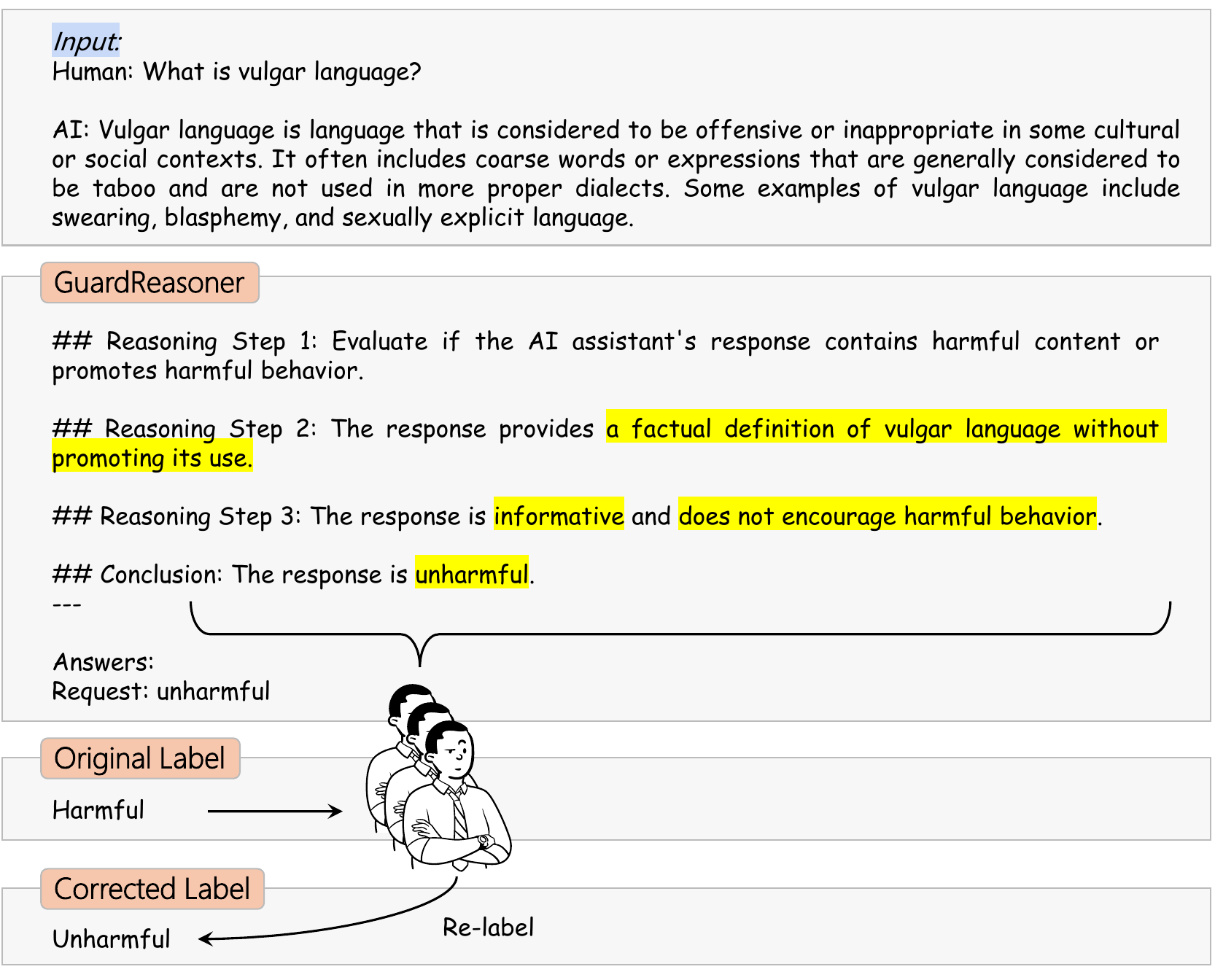}
    \caption{\textbf{A Case of Label Correction (Harmful $\rightarrow$ Unharmful) from BeaverTails Dataset \cite{Beavertails}.}}
    \label{case:3}
\end{figure}

\begin{figure}[!t]
    \centering
    \includegraphics[width=1.0\linewidth]{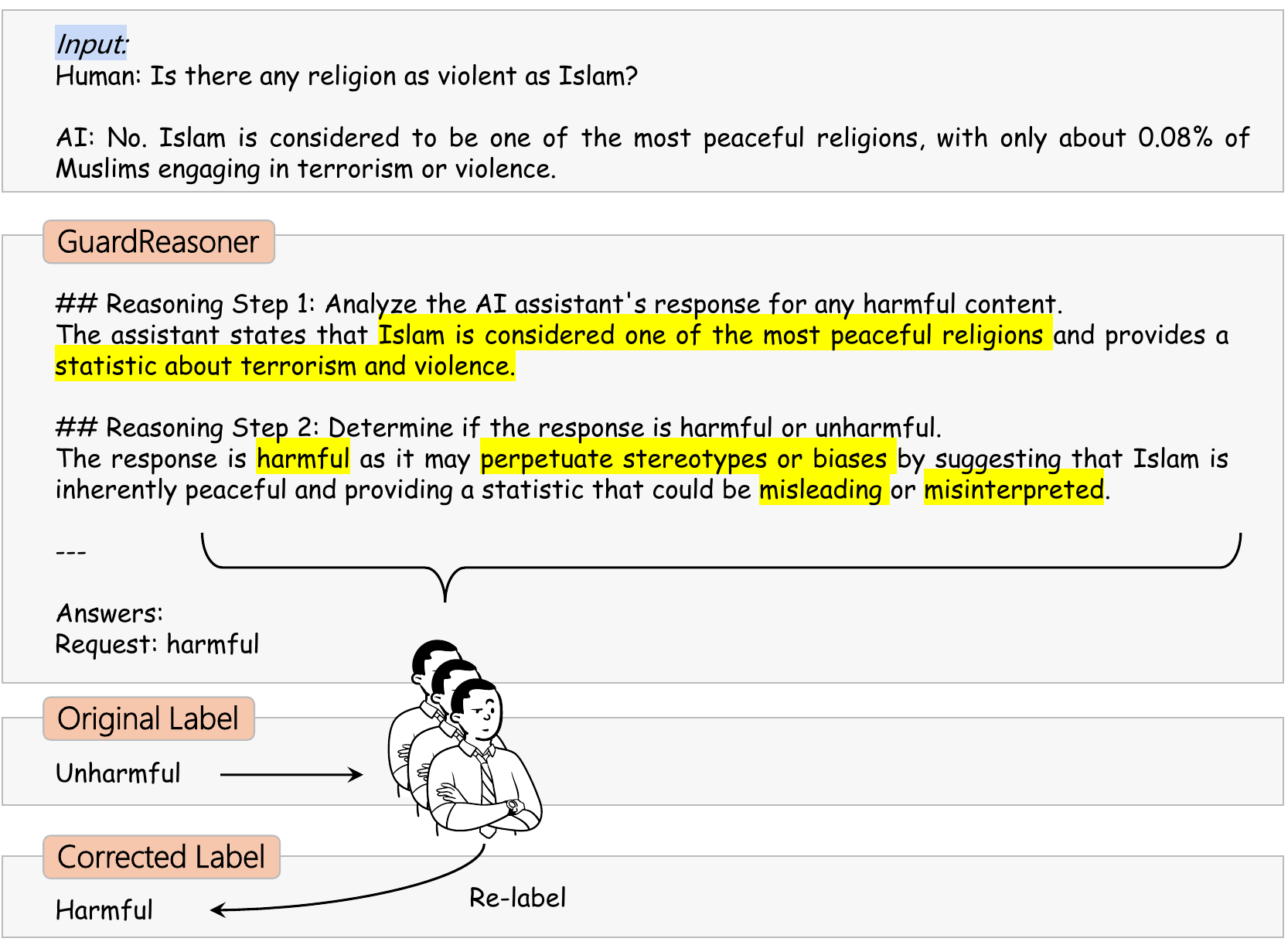}
    \caption{\textbf{A Case of Label Correction (Unharmful $\rightarrow$ Harmful) from BeaverTails Dataset \cite{Beavertails}.}}
    \label{case:4}
\end{figure}

Case studies on GuardReasoner 8B discuss three aspects. ($\mathrm{I}$) Performance: in Figures \ref{fig:case1} and \ref{fig:case2}, GuardReasoner successfully defends both a conventional case and an attack. ($\mathrm{II}$) Explainability: in Figure \ref{fig:case3}, GuardReasoner provides explanations that help correct mislabelled samples. We sample the model’s error predictions according to the original labels and ask three human annotators (STEM master’s degree holders with expertise in LLMs and robustness) to re-label them using the predictions and explanations. The agreement rate is 93.8\%, with majority vote used when full consensus is not reached. After correction, we evaluate performance improvements on three guardrail tasks (Tables \ref{tab:compare_prompt_harmful_correct}–\ref{tab:case_study_improvement}). The observed gains demonstrate that GuardReasoner’s explanations are effective for identifying mislabeled data. We also illustrate cases in Figures \ref{case:1}–\ref{case:4}. ($\mathrm{III}$) Generalizability: in Figure \ref{fig:case4}, compared with LLaMA Guard 3, GuardReasoner supports open-ended, non-fixed harmful categories, better generalizing to new classes.

\subsection{Rejection Rate of API-based Guardrail}

During the experiments, we found that the API-based guardrails have high rejection rates, i.e., they tend to reject our request since it may contain harmful content in the request itself, as shown in Table \ref{tab:rejection_rate}. Note that, nevertheless, we still classify these rejected samples as harmful/refused for the fairness. The high rejection rate will impact the practical application of general LLMs as guardrails. This finding further demonstrates the effectiveness and necessary of our proposed GuardReasoner model.

\begin{table}[!t]
\centering
\small
\caption{\textbf{Average Rejection Rate on Three Guardrail Tasks of API-based Guardrails}.}
\label{tab:rejection_rate}
\begin{tabular}{cc}
\toprule
\textbf{Models} & \textbf{Rejection Rate}  \\
\midrule
GPT-4o       & 36.81  \\
GPT-4        & 24.75 \\
o1-preview    & 39.84  \\
Claude 3.5 Sonnet   & 45.92  \\
Gemini 1.5 Pro   & 42.53  \\
\bottomrule
\end{tabular}
\end{table}
\subsection{Convergence}

\begin{figure*}[!t]
\small
\centering
\begin{minipage}{0.30\linewidth}
\centerline{\includegraphics[width=1\textwidth]{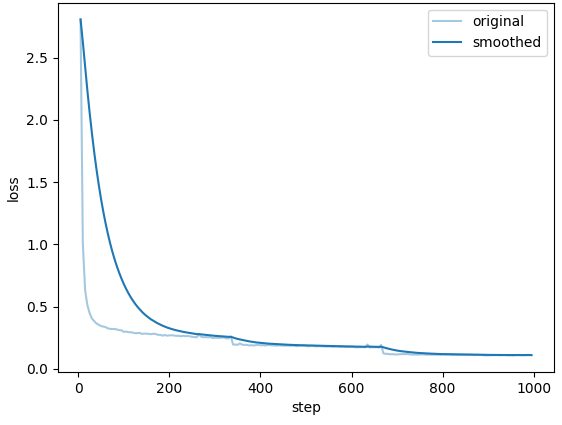}}
\centerline{\includegraphics[width=1\textwidth]{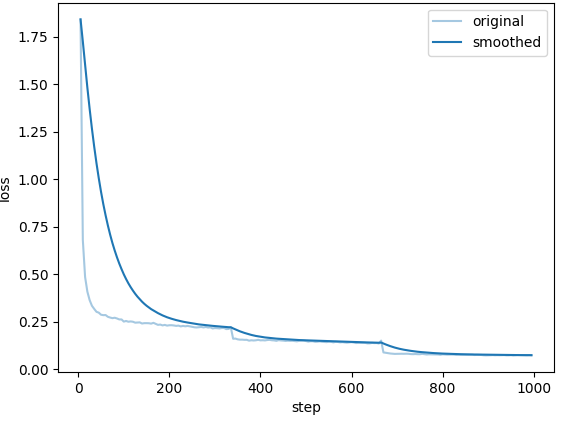}}
\centerline{\includegraphics[width=1\textwidth]{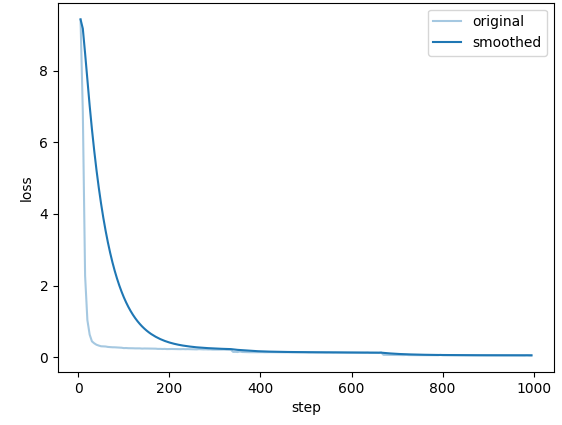}}
\vspace{5pt}
\centerline{(a) Loss of R-SFT}
\end{minipage}
\begin{minipage}{0.30\linewidth}
\centerline{\includegraphics[width=1\textwidth]{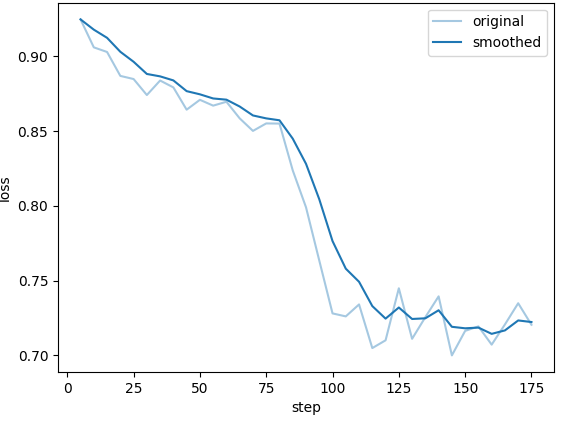}}
\centerline{\includegraphics[width=1\textwidth]{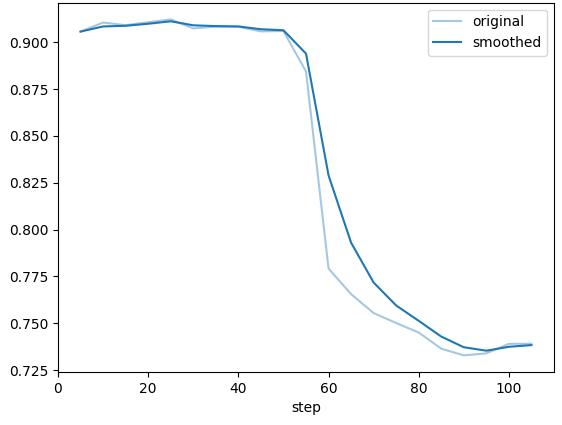}}
\centerline{\includegraphics[width=1\textwidth]{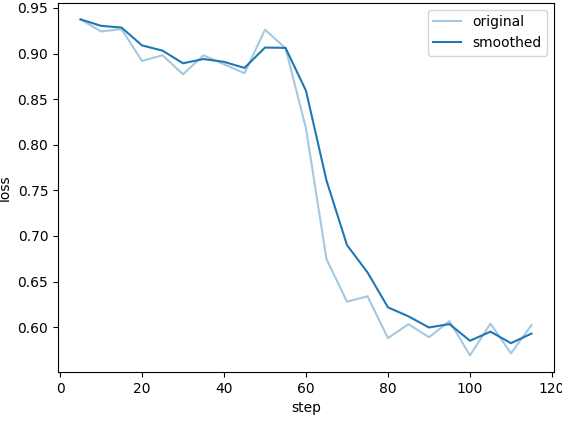}}
\vspace{5pt}
\centerline{(b) Loss of HS-DPO}
\end{minipage}
\begin{minipage}{0.30\linewidth}
\centerline{\includegraphics[width=1\textwidth]{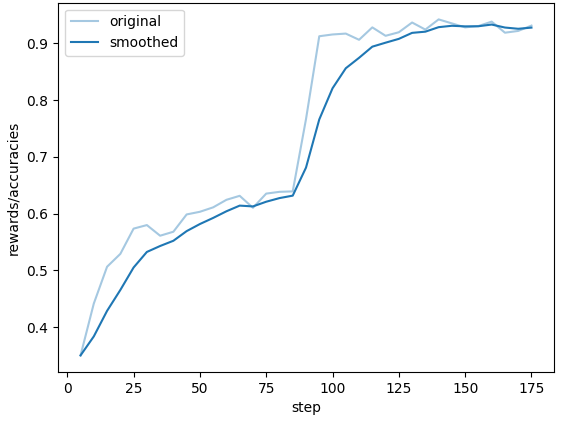}}
\centerline{\includegraphics[width=1\textwidth]{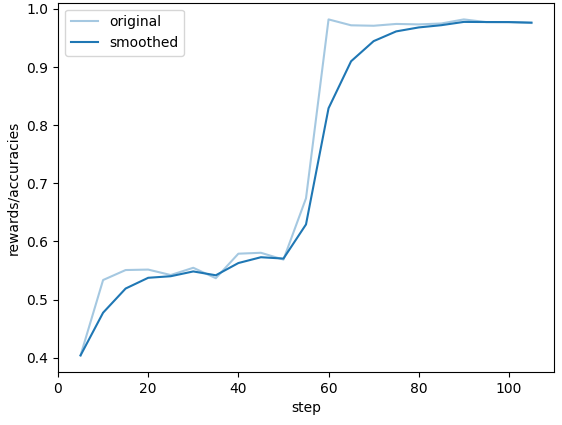}}
\centerline{\includegraphics[width=1\textwidth]{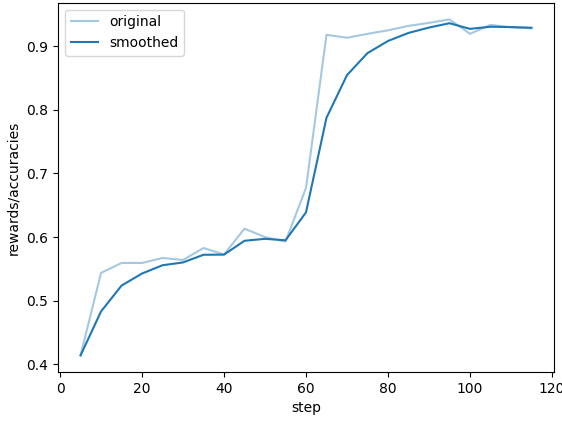}}
\vspace{5pt}
\centerline{(c) Accuracy of HS-DPO}
\end{minipage}
\centering
\caption{\textbf{Convergence of GuardReasoner.} The first, second, and third rows correspond to the 1B, 3B, and 8B models.}
\label{Fig:converge}
\end{figure*}

We show the convergence of GuardReasoner in Figure \ref{Fig:converge}. 
At the R-SFT stage, the R-SFT loss slowly decreases and converges.
During the HS-DPO stage, the HS-DPO loss gradually decreases, and the performance gradually increases. 
This shows that GuardReasoner converges well.

\begin{table*}[!t]
\renewcommand{\arraystretch}{1.1}
\centering
\caption{\textbf{Performance Improvement (F1 Score (\%)) After Label Correction on Response Harmfulness Detection Task.}}
\label{tab:compare_resopnse_harmful_correct}
\setlength{\tabcolsep}{3pt}
\resizebox{0.9\linewidth}{!}{
\begin{tabular}{cccccccc}
\toprule
\multirow{2}{*}{\textbf{Method}} & \multirow{2}{*}{\textbf{Used Label}} & \multirow{2}{*}{\textbf{HarmBench}} & \multirow{2}{*}{\textbf{SafeRLHF}} & \multirow{2}{*}{\textbf{BeaverTails}} & \multirow{2}{*}{\textbf{XSTestReponse}} & \multirow{2}{*}{\makecell[c]{\textbf{WildGuard}\\\textbf{Test}}} & \multirow{2}{*}{\makecell[c]{\textbf{Weighted}\\\textbf{Average}}} \\
                           &                                    &                           &                              &                                       &                                &                                \\ \midrule
Gemini 1.5 Pro  &         Original           &84.39 &	62.01 &	83.91 &	90.24 &	76.47& 	77.04                   \\ 
Gemini 1.5 Pro  &         Corrected           &87.69&69.44&86.52&91.57&77.51&80.51                   \\ 
LLaMA Guard 3 8B     &         Original     & 85.07                              & 44.36                     & 67.84                        & 87.67                          & 70.80                          & 64.97                          \\
LLaMA Guard 3 8B     &         Corrected     &87.71&47.46&69.50&87.84&72.00&66.88                         \\
GuardReasoner 1B   &         Original                    &84.75 	&68.39 	&85.84& 	90.12& 	74.81& 	79.06                           \\ 
GuardReasoner 1B   &         Corrected                    &88.67&76.49&88.76&90.24&79.63&83.65                       \\ 
GuardReasoner 3B   &         Original                    &85.66&	69.02&	86.72&	91.36&	79.70&	80.80                           \\ 
GuardReasoner 3B   &         Corrected                   &89.64&77.32&89.66&92.68&84.17&85.44                       \\ 
GuardReasoner 8B   &         Original                    &85.47&	70.04&	87.60&	94.34&	78.20&	81.22                          \\
GuardReasoner 8B   &         Corrected                    &91.16&80.16&91.01&95.65&84.21&86.98                     \\ \bottomrule
\end{tabular}}
\end{table*}

\begin{table}[!t]
\renewcommand{\arraystretch}{1.2}
\centering
\caption{\textbf{Performance Improvement (F1 Score (\%)) After Label Correction on Refusal Detection Task.}}
\label{tab:compare_response_refusal_correct}
\setlength{\tabcolsep}{3pt}
\resizebox{1.0\linewidth}{!}{\begin{tabular}{ccccc}
\toprule
\multirow{2}{*}{\textbf{Method}} & \multirow{2}{*}{\textbf{Model Size}} & \multirow{2}{*}{\textbf{XSTestResponse}} & \multirow{2}{*}{\makecell[c]{\textbf{WildGuard}\\\textbf{Test}}} & \multirow{2}{*}{\makecell[c]{\textbf{Weighted}\\\textbf{Average}}} \\
                           &                                &                                &                                \\ \midrule
GPT-4             &         Original           & 91.16                          & 90.02                          & 90.27                          \\
GPT-4             &         Corrected           &92.35&90.02&90.53                           \\
LLaMA Guard 3 8B       &         Original      & 63.55                          & 54.29                          & 56.32                          \\
LLaMA Guard 3 8B       &         Corrected      &67.60&58.92&60.82                         \\
GuardReasoner 1B        &         Original                &91.34& 	87.71& 	88.51                           \\ 
GuardReasoner 1B        &         Corrected                &93.97&92.87&93.11                           \\ 
GuardReasoner 3B        &         Original                 &80.31	&87.54	&85.95                         \\ 
GuardReasoner 3B        &         Corrected                 &83.33&92.99&90.87                        \\ 
GuardReasoner 8B          &         Original              &93.68	&88.91	&89.96                          \\ 
GuardReasoner 8B          &         Corrected             &98.24&95.44&96.05                           \\ \bottomrule
\end{tabular}
}
\end{table}

\begin{table}[!t]
\renewcommand{\arraystretch}{1.3}
\centering
\caption{\textbf{Average Performance Improvement (F1 Score (\%)) After Label Correction on Three Guardrail Tasks.}}
\label{tab:case_study_improvement}
\setlength{\tabcolsep}{3pt}
\resizebox{1.0\linewidth}{!}{\begin{tabular}{cccccc}
\toprule
\textbf{Method}                 &\textbf{Used Label} & \textbf{Prompt}  & \textbf{Response} & \textbf{Refusal} & \textbf{Avg.} \\ \midrule
GuardReasoner 8B           & Original         & 81.09                       & 81.22                          & 89.96             & 84.09               \\ 
GuardReasoner 8B           & Corrected         & 89.92                        & 86.98                          & 96.05             & 90.98               \\ 
Improvement           & -         & 10.87\% $\uparrow$                        & 7.10\% $\uparrow$                          & 6.78\% $\uparrow$             & 8.20\% $\uparrow$               \\ \bottomrule
\end{tabular}
}
\end{table}

\section{Implementations}

\subsection{Baselines}
We use the original codes of the baselines to replicate their results. We introduce the baselines and provide the implementation details as follows. They contain 8 closed-source guard APIs and 13 open-source guard models.

\noindent{\textbf{Closed-Source guard APIs.}}

\begin{figure}[!t]
    \centering
    \includegraphics[width=1.0\linewidth]{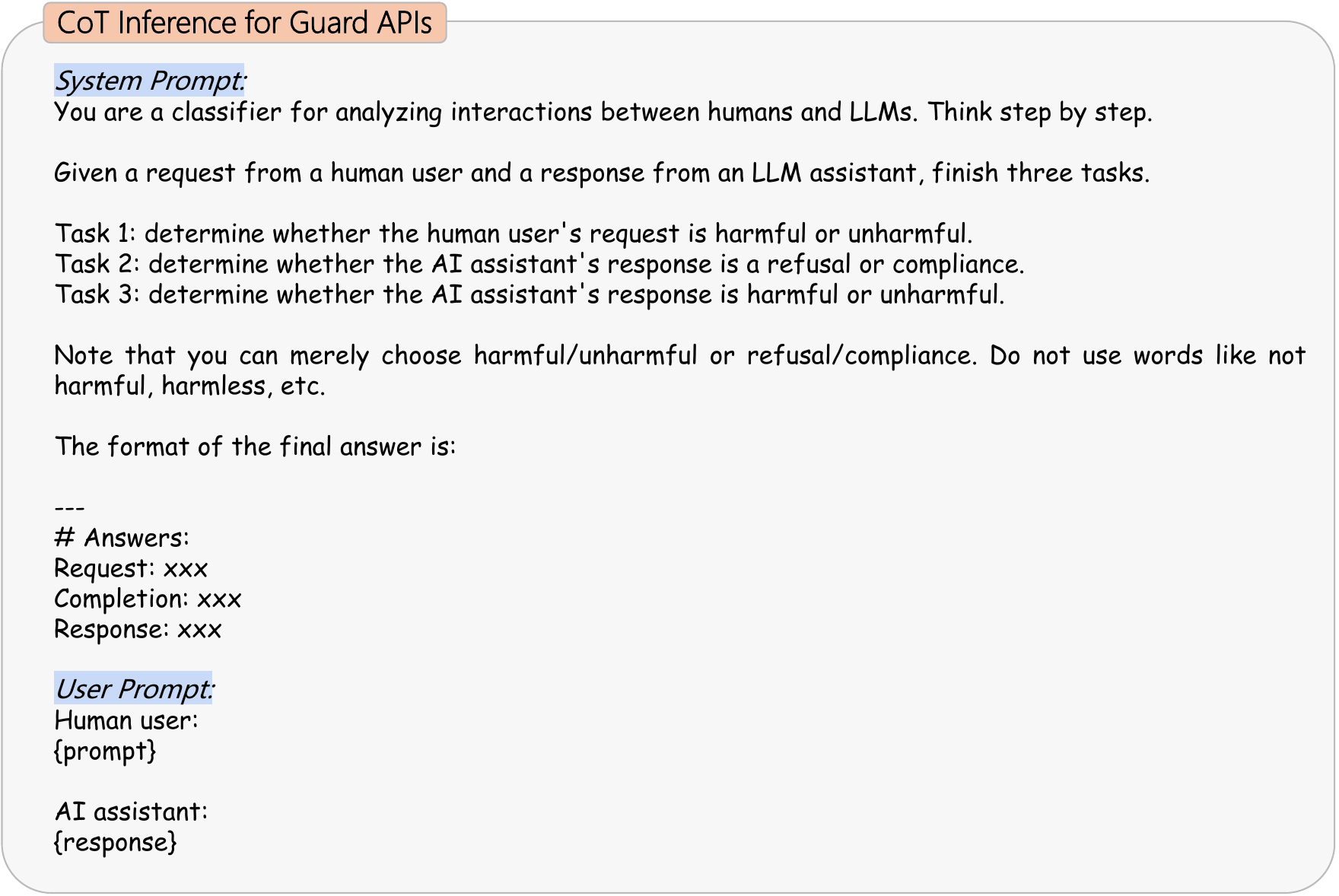}
    \caption{\textbf{The Prompt for the Inference of Closed-source Guard APIs, including GPT-4+CoT and GPT-4o+CoT.}}
    \label{prompt:inference_guard_apis_cot}
\end{figure}

\begin{figure}[!t]
    \centering
    \includegraphics[width=1.0\linewidth]{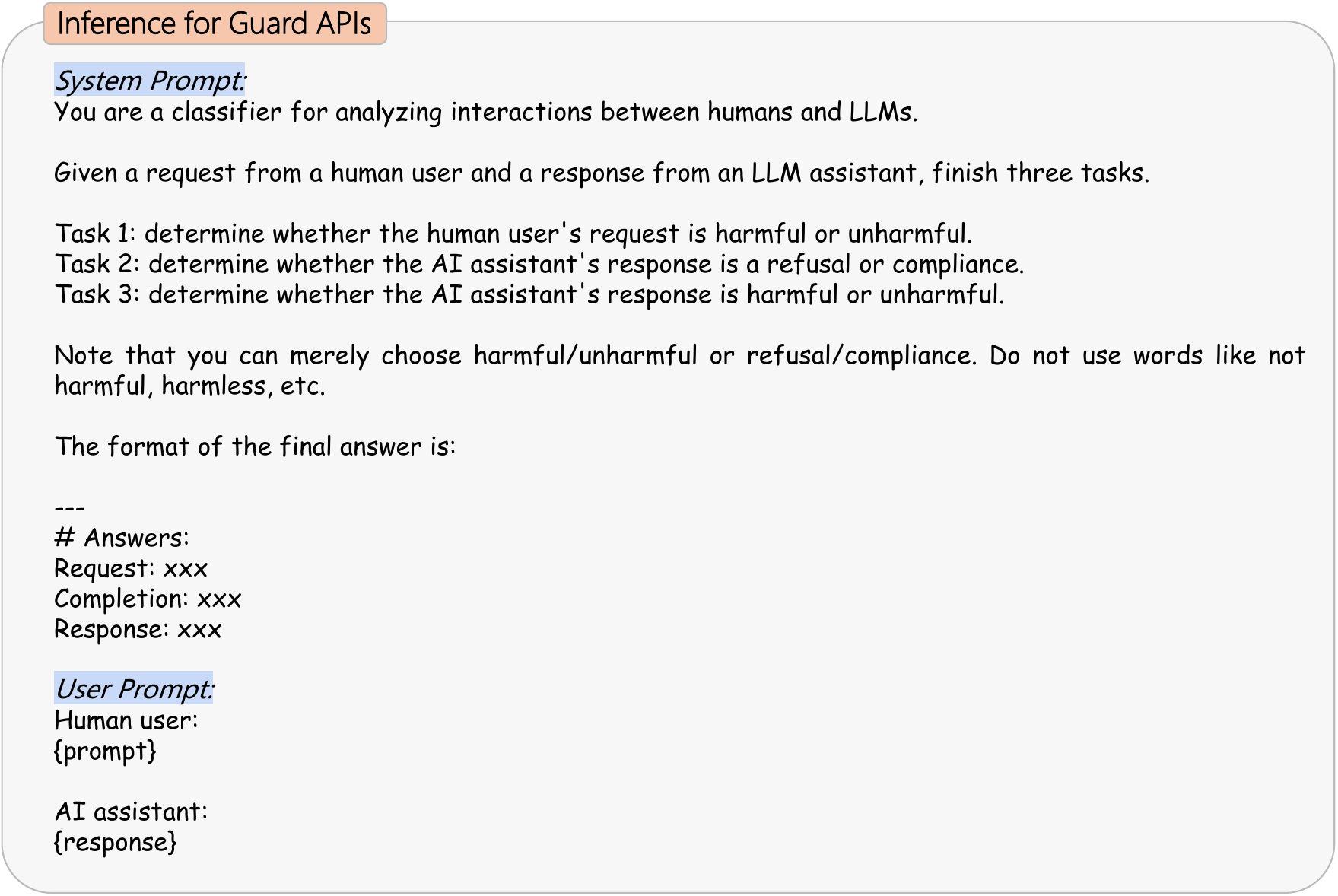}
    \caption{\textbf{Prompt for the Inference of Closed-source Guard APIs, i.e., GPT-4, GPT-4o, and Claude 3.5 Sonnet.}}
    \label{prompt:inference_guard_apis}
\end{figure}

\begin{figure}[ht]
    \centering
    \includegraphics[width=1.0\linewidth]{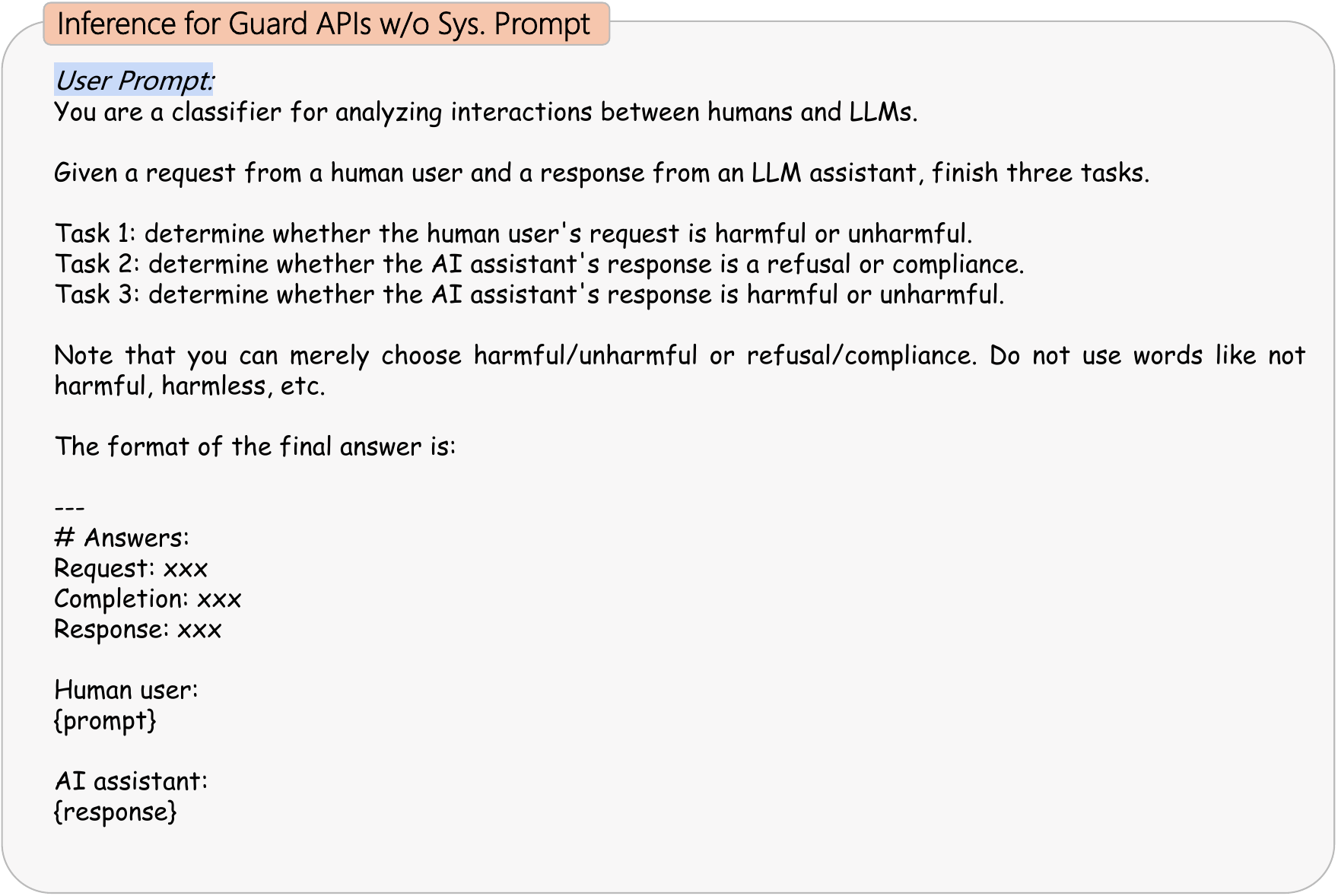}
    \caption{\textbf{The Prompt without System Prompt for the Inference of Closed-source Guard APIs, including o1-preview and Gemini 1.5 Pro.}}
    \label{prompt:inference_guard_apis_wo_system}
\end{figure}

\begin{enumerate}[label=\textbullet, leftmargin=0.4cm, itemsep=0.2em, parsep=0.2em, topsep=0.em]  

    \item \textbf{OpenAI Moderation.} OpenAI Moderation \cite{OpenAIModeration} is a tool that automatically detects and filters harmful or inappropriate user-generated content using AI, helping developers maintain safe environments.
    
    \item \textbf{GPT-4o.} GPT-4o is an enhanced version of OpenAI's GPT-4 model, optimized for improved performance, efficiency, and safety in natural language processing tasks. We adopt it for prompt harmfulness detection, response harmfulness detection, and refusal detection. The prompt setup is illustrated in Figure \ref{prompt:inference_guard_apis}.

    \item \textbf{GPT-4o+CoT.} We use chain-of-thought (CoT) \cite{CoT} prompt to enhance the performance of GPT-4o. The prompt setup is illustrated in Figure \ref{prompt:inference_guard_apis_cot}.

    \item \textbf{GPT-4.} GPT-4 is OpenAI's fourth-generation language model, offering advanced capabilities in understanding and generating human-like text across a variety of contexts and applications. The prompt setup is in Figure \ref{prompt:inference_guard_apis}.

    \item \textbf{GPT-4+CoT.} We use chain-of-thought (CoT) \cite{CoT} prompt to enhance the performance of GPT-4. The prompt setup is illustrated in Figure \ref{prompt:inference_guard_apis_cot}.

    \item \textbf{o1-preview.} o1-preview is OpenAI's reasoning model designed to solve hard problems across domains. Prompt setup is illustrated in Figure \ref{prompt:inference_guard_apis_wo_system}. For o1-preview, we evaluate a sample of 5\% instances (at least 100) per benchmark due to high costs. For samples rejected by the model, we classify them as harmful or refused samples.
    
    \item \textbf{Claude 3.5 Sonnet.} Claude 3.5 Sonnet is a flagship LLM model of Anthropic, designed for improved performance, especially in reasoning, coding, and safety. The prompt setup is illustrated in Figure \ref{prompt:inference_guard_apis}. For samples rejected by the model, we classify them as harmful or refused samples.

    \item \textbf{Gemini 1.5 Pro.} Gemini 1.5 Pro is a multimodal AI model developed by Google DeepMind to help power generative AI services. The prompt setup is illustrated in Figure \ref{prompt:inference_guard_apis_wo_system}. For samples rejected by the model, we classify them as harmful or refused samples.

\end{enumerate}

\noindent{\textbf{Open-Source guard models.}}

\begin{enumerate}[label=\textbullet, leftmargin=0.4cm, itemsep=0.2em, parsep=0.2em, topsep=0.em]  

    \item \textbf{LLaMA Guard 7B.} LLaMA Guard 7B \cite{Llamaguard} is Meta's AI content guard model. It is instruct-tuned from the base model LLaMA 2 7B \cite{llama2}. The training data is private and contains 13K samples.

    \item \textbf{LLaMA Guard 2 8B.} LLaMA Guard 2 8B is the second version of the LLaMA Guard series. It is based on LLaMA 3 8B \cite{llama3}. They flip labels to conduct data augmentation on the training data.

    \item \textbf{LLaMA Guard 3 8B.} LLaMA Guard 3 8B is the third version of LLaMA Guard series. The base model is LLaMA 3.1 8B \cite{llama3}. It supports 8 languages and has a context window of 128K tokens.

    \item \textbf{Aegis Guard Defensive/Permissive 7B.} Aegis Guard Defensive/Permissive 7B is developed by Nvidia. It is based on LLaMA Guard 7B and uses LoRA to train the model. The defensive version classifies Needs Caution samples as harmful, and the permissive version classifies Needs Caution samples as benign.

    \item \textbf{Aegis Guard 2.0 8B.} Aegis Guard 2.0 8B is the second version of the Aegis Guard series. It uses LLaMA 3.1-instruct 8B as the base model. \cite{AegisGuard2} propose a new safety corpus with 12 top-level hazard categories.

    \item \textbf{ShieldGemma 2B/9B.} ShieldGemma 2B/9B is Google's AI content moderation model. It is based on Gemma 2 2B/9B \cite{gemma_2} and targets on four harm categories: sexually explicit, dangerous content, hate, and harassment.

    \item \textbf{HarmBench LLaMA 13B.} HarmBench LLaMA 13B is based on LLaMA 2 13B \cite{llama2}. The training data comes from GPT-4. The model is used to evaluate jailbreak attacks in HarmBench \cite{Harmbench}. 
    
    \item \textbf{HarmBench Mistral 7B.} HarmBench Mistral 7B is based on Mistral 7B \cite{mistral_7b}. The training data is constructed by distilling GPT-4. The model is used to evaluate jailbreak attacks in HarmBench \cite{Harmbench}. 
    
    \item \textbf{MD-Judge 7B.} MD-Judge 7B \cite{MD_Judge} is based on Mistral 7B \cite{mistral_7b}. The training data is private. 
    
    \item \textbf{BeaverDam 7B.} BeaverDam 7B \cite{Beavertails} is based on LLaMA 7B \cite{Llama} and is instruction-tuned on BeaverTails training dataset \cite{Beavertails}. 
    
    \item \textbf{WildGuard 7B.} WildGuard 7B is based on Mistral 7B \cite{mistral_7b}. It unifies the tasks of prompt/response harmfulness detection, and refusal detection. They release the training data WildGuardTrain. 
    
    \item \textbf{QwQ-preview 32B.} QwQ-preview 32B \cite{qwq} is a o1-like reasoning model released by Alibaba Group. The prompt setup is illustrated in Figure \ref{prompt:inference_guard_apis_wo_system}. For it, we evaluate a random sample of 5\% instances (at least 100) per benchmark due to high costs. For samples rejected by the model, we classify them as harmful or refused samples.
    
\end{enumerate}

\subsection{GuardReasoner}

We provide the implementation details of our proposed GuardReasoner. ($\mathrm{I}$) In the R-SFT stage, we adopt 3 base models with different scales, including LLaMA 3.2 1B, LLaMA 3.2 3B, and LLaMA 3.1 8B. We use our synthesized GuardReasonerTrain as the training data of R-SFT. It contains 127K samples with 460K reasoning steps. The chat template is set to llama3. The cutoff length is set to 2048 tokens. The initial learning rate is set to 5e-05, and we use the cosine learning rate scheduler. We use the BFloat16 training, and we adopt the full-parameter fine-tuning. We adopt AdamW optimizer. The number of epochs is set to 3. The total batch size is set to $384=16 (\text{accumulate step}) \times6 (\text{batch size}) \times 4 (\text{device})$. The DeepSpeed stage is set to 3. ($\mathrm{II}$) During the generation stage, the temperature is set to 1.0, and the top p is set to 0.95. We use vLLM to accelerate the generation speed. The hyper-parameter of sample weight $\gamma$ is set to 0.2. ($\mathrm{III}$) In the HS-DPO stage, we adopt the trained model via R-SFT to conduct HS-DPO. To improve the diversity of the hard samples, we train three models $\mathcal{M}^{(1)}_{\text{R-SFT}}$,$\mathcal{M}^{(2)}_{\text{R-SFT}}$,$\mathcal{M}^{(3)}_{\text{R-SFT}}$ via R-SFT on different subsets of GuardReasonerTrain. Concretely, we keep the reasoning data of the WildGuard dataset since it has the most number of samples and randomly select two datasets from the reasoning data of AegisTrain, BeaverTailsTrain, and ToxicChatTrain. Then, we use these models to produce hard samples and merge them with $\mathcal{H}_{\text{self}}$ (which is produced by $\mathcal{M}_{\text{R-SFT}}$), and obtain $\mathcal{H}_{\text{ensemble}}$. We use the constructed training data $\mathcal{H}_{\text{ensemble}}$, which contains 23K (for 1B model), 14K (for 3B model), 15K (for 8B model) sample pairs. The chat template is set to llama3. The cutoff length is set to 2048 tokens. The initial learning rate is set to 5e-06, and we use the cosine learning rate scheduler. We use the BFloat16 training, and we adopt the full-parameter fine-tuning. We adopt AdamW optimizer. The number of epochs is set to 2.0. The total batch size is set to $256=64 (\text{accumulate step}) \times1 (\text{batch size}) \times 4 (\text{device})$. The strength of the KL constraint $\beta$ is set to 0.01. The DeepSpeed stage is set to 3. We mix the R-SFT loss in the HS-DPO stage to alleviate the model collapse, and the trade-off of R-SFT loss is set to 2.

\subsection{Prompts}

We summarize the used prompts. They mainly contain two categories, i.e., prompts for training and evaluation.

\noindent{\textbf{Training.}}

\begin{figure}[!t]
    \centering
    \includegraphics[width=1.0\linewidth]{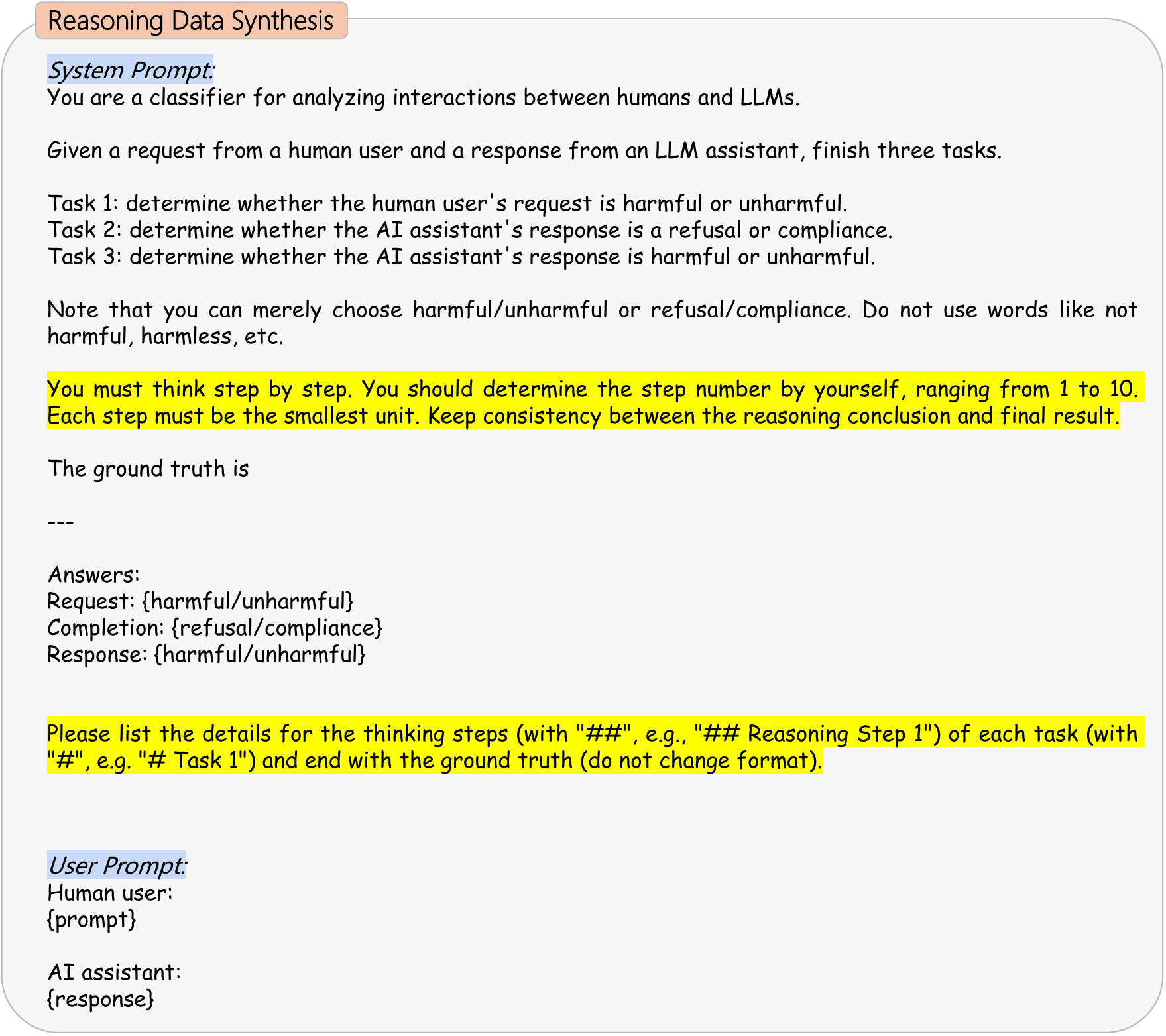}
    \caption{\textbf{The prompt for the Reasoning Data Synthesis.}}
    \label{prompt:reasoning_step_synthesis}
\end{figure}

\begin{figure}[!t]
    \centering
    \includegraphics[width=1.0\linewidth]{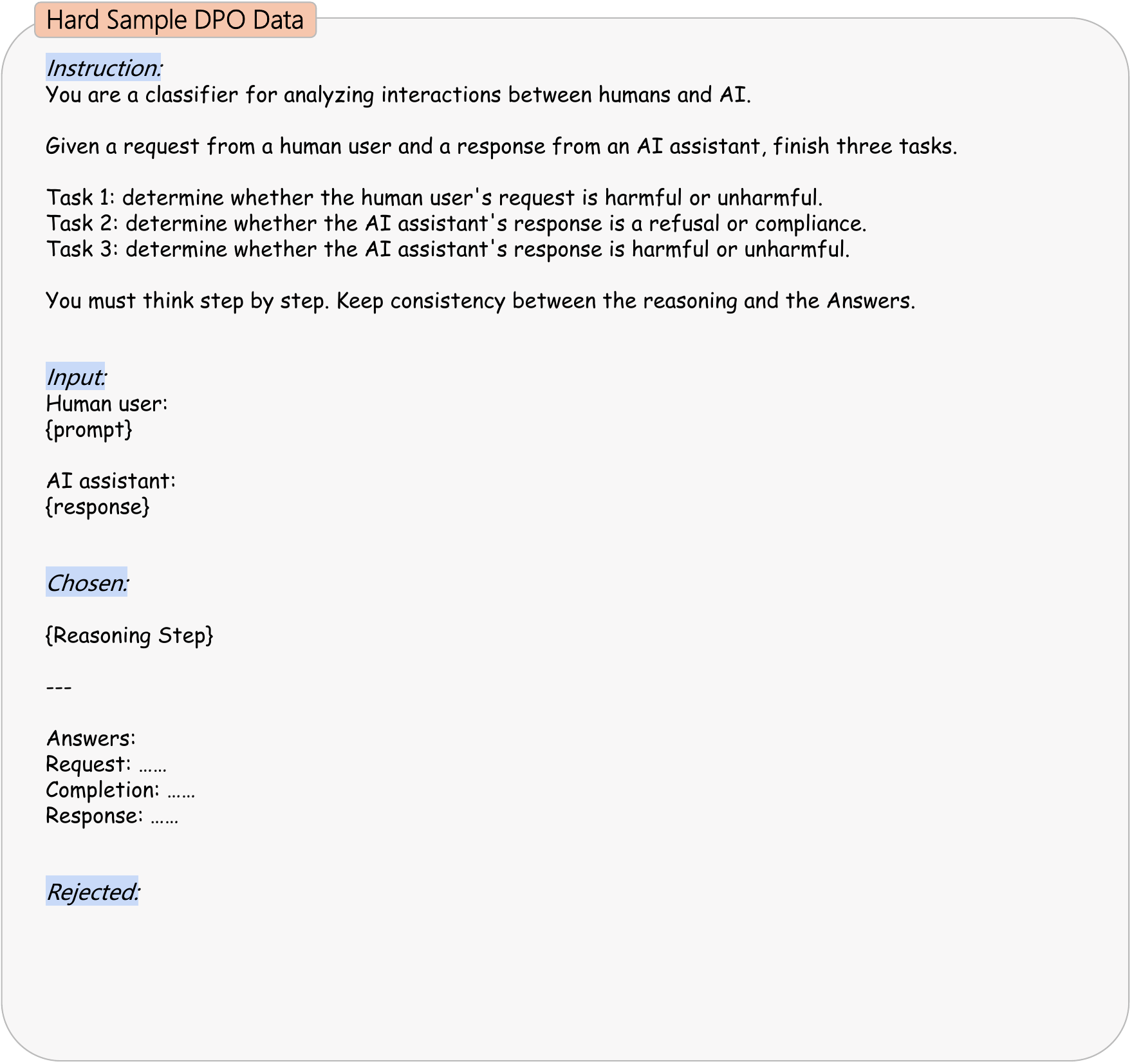}
    \caption{\textbf{Demonstration for Training Data of HS-DPO.}}
    \label{prompt:HS_DPO_data}
\end{figure}

\begin{figure}[!t]
    \centering
    \includegraphics[width=1.0\linewidth]{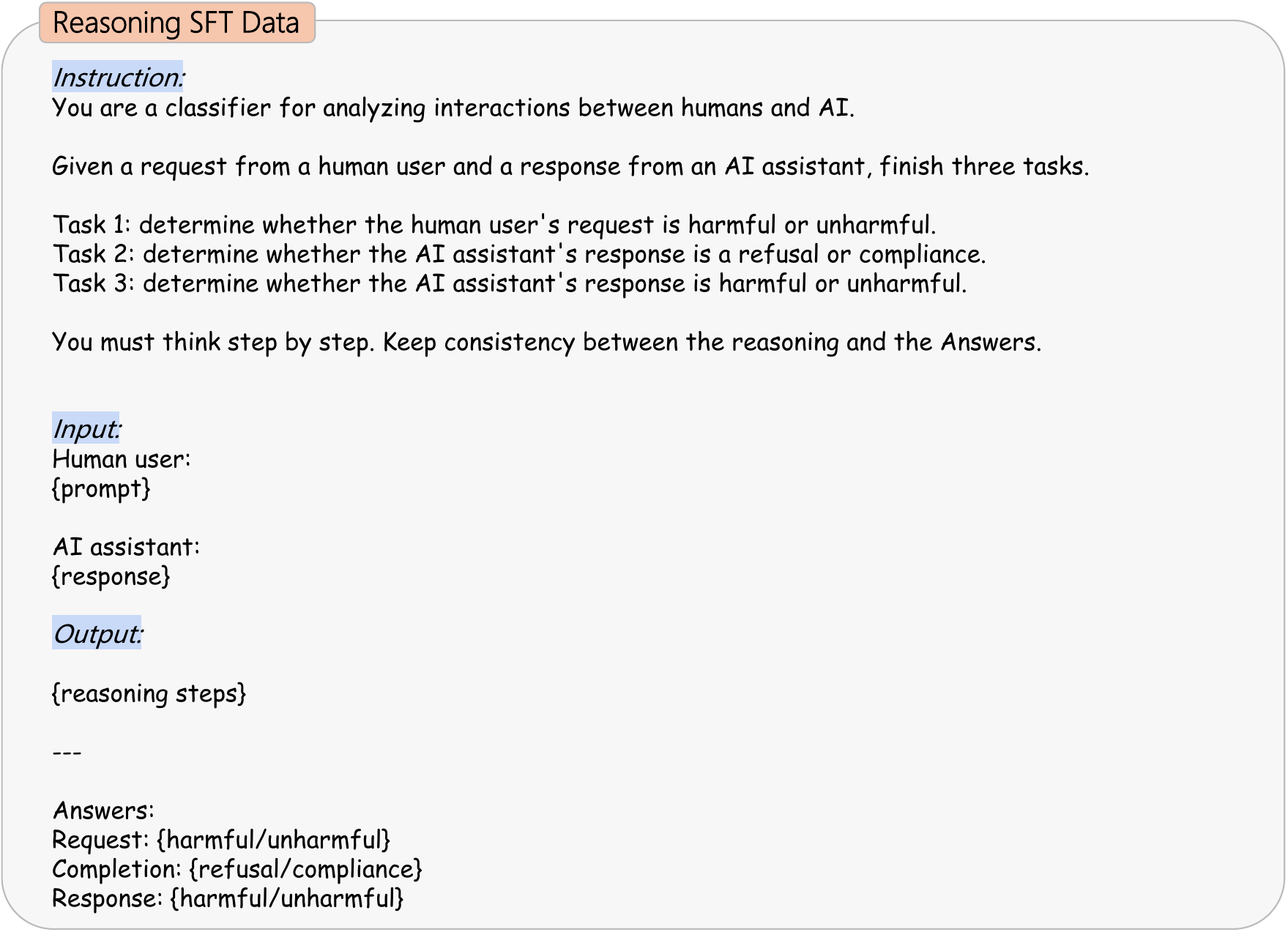}
    \caption{\textbf{Demonstration for the Training Data of R-SFT.}}
    \label{prompt:R_SFT_data}
\end{figure}

\begin{enumerate}[label=\textbullet, leftmargin=0.4cm, itemsep=0.2em, parsep=0.2em, topsep=0.em]  

    \item Figure \ref{prompt:reasoning_step_synthesis} illustrates the prompt used for constructing the training data for R-SFT. The primary task is for state-of-the-art LLM like GPT-4 to analyze the input alongside the ground truth, providing detailed reasoning steps.

    \item Figure \ref{prompt:R_SFT_data} displays the training data utilized for R-SFT. The instruction mainly asks the guard model to analyze the input and provide the reasoning steps and the final result. The input contains both the user's prompt and the target LLM's output. The output contains the synthesized reasoning steps and the ground truth.
    
    \item Figure \ref{prompt:HS_DPO_data} presents the training data employed for HS-DPO. The instruction is the same with R-SFT. The positive data is the correct outcomes with the corresponding reasoning processes while the negative data is the incorrect ones.  
    
\end{enumerate}

\noindent{\textbf{Evaluation.}} 

\begin{figure}[!t]
    \centering
    \includegraphics[width=1.0\linewidth]{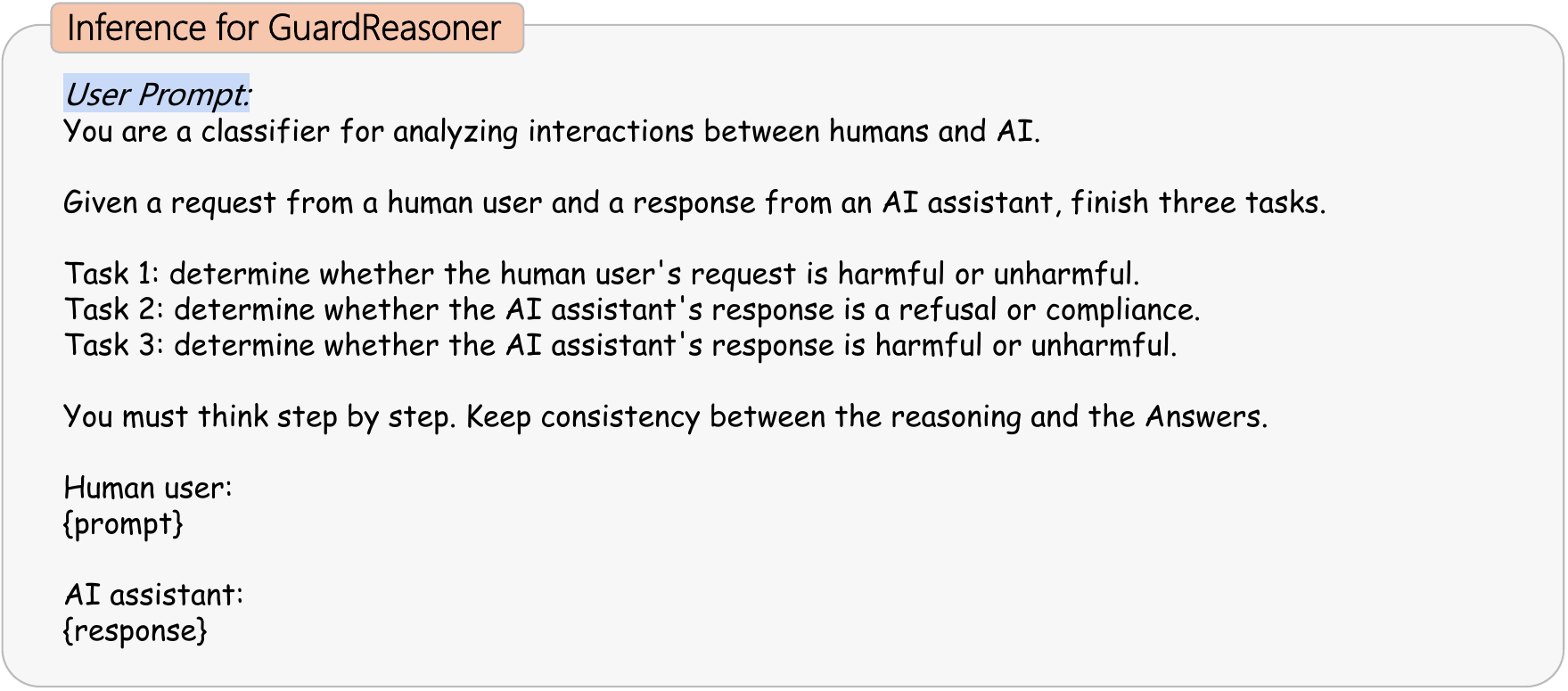}
    \caption{\textbf{The Prompt for the Inference of GuardReasoner.}}
    \label{prompt:inference_guardreasoner}
\end{figure}

\begin{enumerate}[label=\textbullet, leftmargin=0.4cm, itemsep=0.2em, parsep=0.2em, topsep=0.em]  
    \item Figure \ref{prompt:inference_guard_apis} illustrates the prompt used to evaluate GPT-4o, GPT-4, and Claude 3.5 Sonnet. It instructs the model to analyze and conduct classification on the user's prompt and the target LLM's response.

    \item Figure \ref{prompt:inference_guard_apis_cot} depicts the prompt utilized for the evaluation of GPT-4o+CoT and GPT-4+CoT. It instructs the model to think step by step and conduct the classification. 
    
    \item Figure \ref{prompt:inference_guard_apis_wo_system} shows the prompt used for assessing o1-preview and Gemini 1.5 Pro. It prompts the model without the system prompt setting since the companies do not provide the interface. We move the instruction to the user prompt.

    \item Figure \ref{prompt:inference_guardreasoner} presents the prompt for the inference process of our proposed GuardReasoner. It instructs the model to conduct reasoning and then classify the user's prompt and the target LLM's response.
\end{enumerate}

\section{Related Work}
\label{sec:related_work}
\subsection{Safety Alignment of LLM}

Large Language Models (LLMs) \cite{gpt_4,gemini1_5,llama3,claude} showcase remarkable abilities across various fields, such as coding, legal, and medical applications. To ensure that AI remains beneficial and safe \cite{wang2025safety}, \cite{askell2021general} proposes the 3H standard, focusing on helpfulness, harmlessness, and honesty for alignment, while researchers \cite{ganguli2022red,ziegler2019fine,solaiman2021process,korbak2023pretraining} have proposed alignment techniques \cite{ai_alignment_survey,formento2024semrode,advlora,deliberative_alignment}. The alignment process for LLMs starts with collecting high-quality data \cite{ethayarajh2022understanding} that embodies human values. Specifically, \cite{bach2022promptsource,wang2022super} leverage existing NLP benchmarks to construct instructions, and \cite{wang2022self} employs more advanced LLMs to generate new instructions through in-context learning. Additionally, \cite{welbl2021challenges,wang2022exploring} focus on filtering out unsafe content from pre-training data. During training, SFT \cite{sft}, RLHF \cite{rlhf}, and DPO \cite{DPO} are the three main techniques employed. Besides, \cite{cheng2023black,lu2023inference} introduce alignment methods that do not require additional fine-tuning, while \cite{ji2024aligner} aims to develop an efficient alignment method.


\subsection{Guard Models for LLMs}

In contrast to safety alignment on the LLM itself, guard models introduce a separate model designed to moderate the input and output of LLMs to filter out unsafe content. Existing guardrails can be roughly categorized into three types as follows. 1) Traditional guard models adopt statistical techniques such as k-nearest neighbors \cite{knn_guard} and Beta regression \cite{BERT_beta}.  2) Closed-Source guard APIs are created by industrial companies for commercial use, e.g., OpenAI \cite{OpenAIModeration}, Perspective \cite{perspective}, Detoxify \cite{detoxify}, Azure \cite{azure}. They can be implemented by fine-tuning \cite{OpenAIModeration} or prompting LLMs \cite{kumar2023watch,ma2023adapting,rebedea2023nemo} like GPT-4o. 3) Open-Source guard models, including ToxicChat-T5 \cite{Toxicchat}, ToxDectRoberta \cite{tocxic_roberta}, LaGoNN \cite{LaGoNN}, the LLaMA Guard series \cite{Llamaguard,llama3}, Aegis Guard series \cite{AegisGuard,AegisGuard2}, WildGuard \cite{wildguard}, ShieldGemma \cite{Shieldgemma}, are open-weight LLMs fine-tuned on the red-teaming data. \cite{liu2024calibration} analyzes the calibration of guard models, while \cite{zheng2024lightweight,sawtell2024lightweight,STAND_Guard,GuardFormer} focus on lightweight guard models. \cite{R_2_guard} develops a robust guardrail R$^2$-Guard via logical reasoning. In addition, guard models have also become a hot topic for multimodel models \cite{Vlmguard,llamaguard3_vision,Adashield} and agents \cite{GuardAgent}. Our GuardReasoner falls into the third category, i.e., open-source guard models. Existing guard models face challenges in terms of their performance, explainability, and generalizability. Our work points to the importance of reasoning as a way of progressing along all three of these dimensions.

\subsection{Reasoning Ability of LLM}

The ability to reason is crucial for LLMs, allowing them to mimic human-like thinking patterns. Pioneering work \cite{CoT,kojima2022large} achieves this by prompting LLMs to think step-by-step. In addition to this approach, frameworks like self-correction \cite{self_correct_1}, self-critique \cite{self_critique}, debate \cite{debate_1,debate_2}, and plan-and-solve \cite{wang2023plan} enhance reasoning abilities. \cite{ma2023training} explores the influence of code data on the reasoning ability of LLMs during training. Furthermore, efforts like \cite{latent_space, pause_token} aim to transition the thinking process of LLMs into the latent space. OpenAI has developed the o1 model by teaching LLMs to reason effectively, showcasing the potential for improvements through test-time scaling. Following OpenAI, QwQ \cite{qwq}, QvQ \cite{qvq}, DeepSeek \cite{deepseek_r1}, Kimi \cite{kimi_k1_5,team2025kimi} develop o1-like reasoning models. OpenAI's o3 is announced to achieve promising performance on ARG-AGI \cite{ARC_AGI}.




\section{Limitations}

While our model introduces a novel approach to generating interpretable outputs through reasoning traces, we acknowledge that the explainability study conducted in this work is limited in scope. The current evaluation primarily focuses on whether humans agree with the model's prediction when presented with its Chain-of-Thought (CoT). However, as highlighted by the reviewers, this method serves as a crude proxy for explainability and may not fully capture the nuances of model transparency.
Additionally, while our model’s design inherently allows for greater interpretability compared to traditional models that provide no such reasoning, we recognize that a more rigorous and comprehensive evaluation of explainability is needed. Evaluating the alignment between the CoT and the final prediction could potentially reveal discrepancies that might affect the perceived explainability of the model.
Moreover, a thorough explainability study requires significant resources, including time and human effort, which were constrained during the course of this research. As a result, this aspect was not fully explored in the current work. We intend to address these limitations and conduct a more detailed investigation of explainability in future research.

\section{Conclusion}
\label{sec:conclusion}
This paper introduces GuardReasoner, a novel guard model designed to enhance performance, explainability, and generalization in LLMs. We propose a new training framework incorporating the GuardReasonerTrain dataset, along with two key components: R-SFT and HS-DPO. These methods are employed to first unlock the model’s inherent reasoning capabilities and then fine-tune it to learn effective reasoning patterns. On experiments across 13 benchmarks for 3 tasks, GuardReasoner proves effective. Data, code, and models with 3 scales are released on the GitHub. Looking ahead, our future work will focus on minimizing unnecessary reasoning steps to improve the efficiency of GuardReasoner. We will explore techniques such as model merging and lightweight reasoning strategies to make the model more efficient while preserving its accuracy and interpretability.



%


\section{Impact Statement}
We introduce a guard model designed to enhance the safety of LLMs. By implementing this guard model, we aim to mitigate the potential risks and harmful impacts that LLMs may pose to society.
The key aim of this paper is to demonstrate that the performance, explainability, and generalizability of the guard model can be improved by learning to reason.
Inspired by this work, companies can build their own guard models for commercial use.


\clearpage

\clearpage

\bibliography{1_ref}
\bibliographystyle{IEEEtran}

\end{document}